\documentclass[fleqn,usenatbib]{mnras}

\usepackage{newtxtext,newtxmath}

\usepackage[T1]{fontenc}

\DeclareRobustCommand{\VAN}[3]{#2}
\let\VANthebibliography\thebibliography
\def\thebibliography{\DeclareRobustCommand{\VAN}[3]{##3}\VANthebibliography}

\usepackage{graphicx}	
\usepackage{amsmath}	
\usepackage[normalem]{ulem}

\usepackage{xcolor}

\newcommand{\tsup}{\, {\tau_{\rm s} }} 
\newcommand{\tacc}{\, {\tau_{\rm acc} }} 
\newcommand{\name}[1]{{{\it #1}}}

\title[Superradiance for variable accretion]{The impact of superradiance on the spin evolution of variably accreting massive black holes}

\author[Nandakumar et al.]{
Adithya Nandakumar,$^{1}$
Ricarda S. Beckmann,$^{2,4}$\thanks{E-mail: ricarda.beckmann@roe.ac.uk (RSB)}
and Vid Ir\v{s}i\v{c}$^{3,4}$\thanks{E-mail: v.irsic@herts.ac.uk (VI)}
\\
$^{1}$Intitute of Astronomy, University of Cambridge, Madingley Road, Cambridge CB3 0HA, UK\\
$^{2}$Institute for Astronomy, University of Edinburgh, Royal Observatory, Edinburgh EH9 3HJ, UK\\
$^{3}$Center for Astrophysics Research, Department of Physics, Astronomy and Mathematics, University of Hertfordshire, College Lane, Hatfield AL10 9AB, UK\\
$^{4}$Kavli Institute for Cosmology, University of Cambridge, Madingley Road, Cambridge CB3 0HA, UK
}

\date{\rm Accepted 2026 January 19. Received 2025 December 29; in original form 2025 October 24}

\pubyear{2015}

\begin{document}
\label{firstpage}
\pagerange{\pageref{firstpage}--\pageref{lastpage}}
\maketitle

\begin{abstract}
This paper explores how time-varying increases in mass accretion onto rapidly spinning black holes influence their long-term spin evolution when affected by superradiance - a process where energy is extracted from the black hole by a surrounding axion field. Using simulations the study tracks how sudden accretion boosts affect a critical spin-down phase (the superradiance drop) during which the black hole’s spin rapidly decreases while its mass remains nearly constant. The black hole spin evolution is controlled by the competition between two  processes: how fast angular momentum is added through accretion, and how fast it is removed by the axion cloud. One major conclusion is that boosts to the accretion rate {\it before the superradiance drop} have the strongest effect, as they can delay or reshape the drop and significantly shrink the region of the mass-spin plane depopulated due to the superradiance. In particular, a super-Eddington accretion rate of 5 times Eddington accretion, lasting for 4 Myr and occurring 30 Myr before the superradiance drop can reduce the superradiance exclusion region in the mass-spin plane by 40 \%. In contrast, boosts to the accretion rate {\it after the superradiance drop} only cause temporary changes in the black hole spin. The study also shows that black holes with lighter axion clouds are more sensitive to these early boosts and can show observable spin changes lasting tens to hundreds of millions of years. Heavier axion clouds, however, require much stronger or longer-lasting boosts to produce similar effects, making them more stable under variable accretion.
\end{abstract}




\section{Introduction}
\label{sec:introduction}
For light bosonic fields (axions specifically), hydrogen-like gravitational orbitals form around the black hole, much like the electromagnetic orbitals of electrons in hydrogen atoms \citep{Baumann1,arvanitaki2011ExploringStringAxiverse,PhysRevD.95.043001}. These orbitals are known as axion clouds. The orbitals are derived by solving the Klein-Gordon equations for a scalar field in the Kerr metric. When the black hole is spinning fast enough (such that its horizon angular velocity is greater than the angular phase velocity of a given orbital wave mode) we observe a non-classical transfer of angular momentum and energy from the black hole to the orbital \citep{zeldovich,Starobinsky,bardeen,PhysRevD.22.2323}. This effect, known as superradiance, continues until the black hole spins down enough so that the superradiance condition is no longer met. This leads to certain mass/spin values of black holes being disfavoured as the superradiance effect should rapidly evolve black holes away from specific regions in the mass-spin parameter space (called the Regge plane) \citep{arvanitaki2011ExploringStringAxiverse}. As the orbital phase velocity depends on the mass of the boson, we can use data from the observed masses and spins of black holes to constrain the mass of possible bosonic fields \citep{brito,paper1}. 

Observational signatures of superradiance have been explored using the measurements of stellar black holes to exclude the mass of axions ($\mu$) in the range of $6\times10^{-13}eV<\mu<10^{-12}eV$ \citep{Cardoso_2018}, and supermassive black holes to exclude the mass of the axions in the range of $7\times10^{-20}eV<\mu<10^{-16}eV$ \citep{paper1}. Further work has been done on the theoretical front by solving these orbitals at higher orders, including self-interactions, relativistic effects and coupling up to $n=5$ \citep{Witte:2024drg,TaillteMay,sarmah2025EffectsSuperradianceActive,PhysRevD.103.095019,PhysRevD.87.043513}. Superradiance effects have also been analysed on more complex black holes such as charged black holes \citep{chargedbh}, which showed that electromagnetic scattering intensity can be significantly enhanced by superradiance. The emission of the gravitational waves from the black hole and bosonic cloud system has also been predicted \citep{arvanitaki2011ExploringStringAxiverse,PhysRevD.95.043001,Saha_2022,Davoudiasl_2019}. Similarly to the atomic case, binary black holes experience interaction between both orbitals that can imprint themselves on emitted gravitational waves \citep{PhysRevD.100.044051,Baumann2}. Binary interactions themselves can also destabilise the boson cloud and even terminate superradiance \citep{Fan:2023jjj}. 


However, black holes do not exist in isolation in nature. Therefore, matter surrounding the black hole can accrete onto it and affect the superradiance process. Several studies have explored the effect of dark matter accretion \citep{Hui} and baroynic accretion \citep{paper1,brito} under the assumption of a constant matter accretion rate onto the black hole. The main conclusions for baryonic accretion showed that accretion can drive a black hole to a stronger superradiance effect by increasing the black hole mass, which in turn reduces the superradiance extraction timescale. This allows a black hole, which previously had a low enough mass to be far away from the exclusion region, to be driven towards it and hence undergo rapid superradiant extraction. \citep{brito} further showed that the energy density of the built-up scalar cloud is very low even though the scalar cloud becomes a sizeable fraction of the black hole mass. This is because the cloud is spread over a large distance, with typical densities of the cloud below $10^{-8}$ of the BH density\citep{brito} $\rho_{\rm BH}$\footnote{The black hole density is $\rho_{\rm BH} = \frac{3c^6}{32\pi G^3M_{BH}^2}$}
Therefore, the backreaction is negligible, and the Kerr metric still accurately describes the black hole. For this reason, we will ignore backreaction effects and perturbations from the Kerr metric throughout this project. 

It is well established that massive black hole growth is not constant across cosmic time. Instead, the bulk of black hole mass is assembled through multiple periods of radiatively efficient accretion with overall lifetimes on the order of $10^8 \rm yr$ or more \citep{sotan1982MassesQuasars,yu2002ObservationalConstraintsGrowth,mclure2004CosmologicalEvolutionQuasar,trakhtenbrot2011BLACKHOLEMASS}. Simulations have shown that even during active phases, accretion rates onto black holes are highly variable. The clumpy nature of the interstellar medium naturally leads to variability in the feeding rate of black holes on timescales below a Myr ($10^6 \rm yr$) \citep[e.g.][]{dubois2015BlackHoleEvolution,angles-alcazar2013BLACKHOLEGALAXYCORRELATIONS,Ricarda,prieto2017HowAGNSN}. More sustained boosts in accretion rate, lasting $10^7$ years or more, can be triggered by galaxy-wide events such as galaxy mergers \citep[][]{hopkins2006UnifiedMergerdrivenModel,dimatteo2008DirectCosmologicalSimulations,mcalpine2018RapidGrowthPhase,lapiner2021CompactiondrivenBlackHole}. In the early Universe, such accretion bursts can exceed the Eddington limit  \citep{regan2019SuperEddingtonAccretionFeedback,massonneau2023SuperEddingtonGrowth,husko2025EffectsSuperEddingtonAccretion,lupi2024SustainedSuperEddingtonAccretion}.

In this paper we investigate how periods of efficient accretion affect the distribution of black hole spins and masses for supermassive black holes in the presence of a scalar cloud. The remainder of this paper is organised as follows:  Sec. \ref{sec:model} and Sec. \ref{sec:params_names} detail how the black hole superradiance model is created and any nomenclature we define. Sec. \ref{sec:timescale_intro} develops a framework for analysing the timescales to understand how superradiance works. Then low boost and high boost variable accretion models are added to the black hole simulation in Sec. \ref{sec:weak_boost} and Sec. \ref{sec:strong_boost} respectively. The axion mass is then varied and the effects of variable accretion are investigated in Sec. \ref{sec:varying_ax_mass}. The effects of the accretion events are then collated and analysed against the exclusion region in Sec. \ref{sec:exclusion_area}.

\section{Black hole superradiance with accretion events}
\subsection{Superradiance Model}
\label{sec:model}
In this paper we use the model from \citet{brito}, to model the impact of black hole superradiance on black hole spin in the presence of accretion. Adjusting the model for our unit convention gives the following equations:

\begin{equation}
    \omega_{\rm I}=\frac{1}{48}\frac{c^4}{GM}\left( \frac{J}{GM^2}-\frac{2\mu r_+}{\hbar}\right) \alpha^9.
    \label{eq:freq}
\end{equation}
\begin{equation}
    \dot{M}_s=2M_{\rm S}\omega_{\rm I},
    \label{eq:dotMcloud}
\end{equation}
\begin{equation}
    \dot{M}=\dot{M}_{\rm ACC}-\dot{M}_s,
    \label{eq:Massconserveequation}
\end{equation}
\begin{equation}
    \dot{J}=\dot{J}_{\rm ACC}-\frac{\hbar\dot{M}_s}{\mu}.
    \label{eq:AMconserveequation}
\end{equation}
where $\alpha = \frac{G\mu M}{c \hbar}$. $M_{\rm S},M$ and $\mu$ correspond to the axion cloud mass, black hole mass and axion mass respectively. $\omega_{\rm I}$ is the frequency term for the lowest energy orbital that exhibits superradiance and can be understood as the timescale of growth for a particular orbital. $\dot{M}_{\rm ACC}$ and $\dot{J}_{\rm ACC}$ are the rate of changes of mass and angular momentum, due to the external baryonic accretion into the black hole. $\hbar$ 
 is the Planck constant, $G$ 
 the gravitational constant and $c$ 
 the speed of light. $r_+$ is the event horizon for a rotating black hole given by $r_+=\frac{GM}{c^2}+\sqrt{\frac{G^2M^2}{c^4}-\frac{J^2}{G^2M^2c^2}}$. Equations \ref{eq:Massconserveequation} and \ref{eq:AMconserveequation} are conservation equations for mass and angular momentum respectively.

The lowest energy orbital that exhibits superradiance creates the strongest superradiance effect and dominates the shape of the exclusion region. The superradiant extraction timescale for the next orbital exceeds the age of the universe for supermassive black holes. For this reason, only the lowest energy orbital is considered here.

The dimensionless black hole spin parameter is defined as $a=\frac{cJ}{GM^2}$. When $\omega_{\rm I} \rightarrow 0$ and no more angular momentum is exchanged between axion cloud and black hole, $a \rightarrow a_{\rm crit}$ where:
\begin{equation}
    a_{\rm crit}=\frac{4\alpha}{1+4\alpha^2}.
    \label{eq:acrit}
\end{equation}
This allows us to rewrite $\omega_{\rm I}$ as:
\begin{equation}
    \omega_{\rm I}=\frac{1}{48}\frac{c^3}{GM}(a-a_{\rm crit})\alpha^9.
    \label{eq:freqspin}
\end{equation}

To model the rate of gas mass accretion $\dot{M}_{\rm ACC}$ onto the black hole, we assume a thin accretion disc and parameterise $\dot{M}_{\rm ACC} = f_{\rm Edd}\dot{M}_{\rm Edd}$ where $f_{\rm Edd}$ is the Eddington ratio and $\dot{M}_{\rm Edd}$ is the Eddington mass accretion limit. To compute the angular momentum accretion onto the black hole we follow \citet{bardeenKerrMetricBlack1970}:
\begin{equation}
    \dot{J}_{\rm ACC}=\frac{2GM}{3c\sqrt{3}}\frac{1+2\sqrt{\frac{3c^{2}r_{\rm ISCO}}{GM}-2}}{\sqrt{1-\frac{2GM}{3c^{2}r_{\rm ISCO}}}}\dot{M}_{\rm ACC},\label{eq:Jaccterm}
\end{equation}
where $r_{\rm ISCO}$ is the ISCO radius of the Kerr black hole. The positive sign of the angular momentum accretion term also implies we are only investigating the pro-grade accretion events.

\subsection{Model parameters and nomenclature}
\label{sec:params_names}
We visualise the impact of superradiance by plotting the evolution of our black holes on the black hole mass - spin plane, which is known as the Regge plane. The exclusion region is defined to be the region of the Regge plane predicted to be devoid of black holes due to superradiance effects. The exclusion region is an important diagnostic tool to constrain axion masses by comparing predicted exclusion regions to observed black hole mass and spin values. 

In this paper, we investigate the impact of discrete mass accretion events on the distribution of black holes in the Regge plane, with a particular view to understanding their impact on the shape of the exclusion region. Each discontinuous boosts in accretion is applied at a fixed Eddington ratio $f_{\rm Edd}$, starting at a time $t_{\rm ev}$ and lasting for a duration of $\Delta t_{\rm ev}$. 
Simulations presented in this paper are labelled as follows:
\begin{enumerate}
\setlength\itemsep{0.5em}
    \item By the strength of their accretion boost: L(ow) for a boost to $f_{\rm Edd}= 0.5$, H(igh) for a boost to $f_{\rm Edd}= 5$.
    \item By the starting time of the boost $t_{\rm ev}$: E(arly) for $t_{\rm ev} = 0.16 \rm Gyr$, M(id) for $t_{\rm ev} = 0.26 \rm Gyr$ and L(ate) for $t_{\rm ev} = 0.36 \rm Gyr$
    \item A number to denote the boost duration $\Delta t_{\rm ev}$ in Myr.
\end{enumerate}

Simulations with boosted accretion are compared to a fiducial simulation that retains $f_{\rm Edd} = 0.05$ throughout. For example, a boost to $f_{\rm Edd} = 5$ at $t_{\rm ev} = 0.26 \rm \ Gyr$ for 1 Myr would be denoted as HM1.

\begin{table}
\centering
\renewcommand{\arraystretch}{1.2}  
\begin{tabular}{|p{2.8cm}|p{0.5cm}|p{0.5cm}|p{0.5cm}|p{0.5cm}|p{0.5cm}|p{0.5cm}|}
\hline
\textbf{$t_{ev}$(Gyr) $\backslash$ $\Delta t_{ev}$(Myr)} & 10 & 15 & 20 & 25 & 30 & 40 \\
\hline
0.16  & LE10 & LE15 & LE20 & LE25 & LE30 & LE40 \\
\hline
0.26  & LM10 & LM15 & LM20 & LM25 & LM30 & LM40 \\
\hline
0.36  & LL10 & LL15 & LL20 & LL25 & LL30 & LL40 \\
\hline
\end{tabular}
\caption{Model labels for boosts to $f_{edd}=0.5$. The columns represent $\Delta t_{ev}$ which is the duration of the accretion boosts. The rows correspond to $t_{ev}$, which is the time when the boosts are applied. Each cell corresponds to a specific simulation scheme.}
\label{table:0.05sim}
\end{table}

\begin{table}
\centering
\renewcommand{\arraystretch}{1.2}  
\begin{tabular}{|p{2.8cm}|p{0.5cm}|p{0.5cm}|p{0.5cm}|p{0.5cm}|p{0.5cm}|p{0.5cm}|}
\hline
\textbf{$t_{ev}$(Gyr) $\backslash$ $\Delta t_{ev}$(Myr)} & 1 & 1.5 & 2 & 2.5 & 3 & 4 \\
\hline
0.16  & HE1 & HE1.5 & HE2 & HE2.5 & HE3 & HE4 \\
\hline
0.26  & HM1 & HM1.5 & HM2 & HM2.5 & HM3 & HM4 \\
\hline
0.36  & HL1 & HL1.5 & HL2 & HL2.5 & HL3 & HL4 \\
\hline
\end{tabular}
\caption{Model labels for boosts to $f_{edd}=5$. The structure of the table is the same as table \ref{table:0.05sim}.}
\end{table}

In this work, we ignore relativistic corrections when the black hole spin approaches unity, and therefore limit the maximum allowed spin to be below $a=0.998$ \citep{kip}. In practice, we set $\dot{J}_{\rm ACC}=0$ when $a=0.998$ and only include a non-zero accretion term in the angular momentum conservation equations when $a\leq0.997$. Mass accretion $\dot M_{\rm ACC}$ remains constant throughout. To predict the impact of each accretion event on the Regge plane, we numerically solve the equations laid out in Sec. \ref{sec:model} for a range of initial conditions that vary:
\begin{itemize}
    \item The initial seed mass of axion cloud $M_0$
    \item The axion mass $\mu$
    \item The initial black hole mass $M$
    \item The initial spin of black hole $a$
\end{itemize}

\begin{figure}
	\includegraphics[width=\columnwidth]{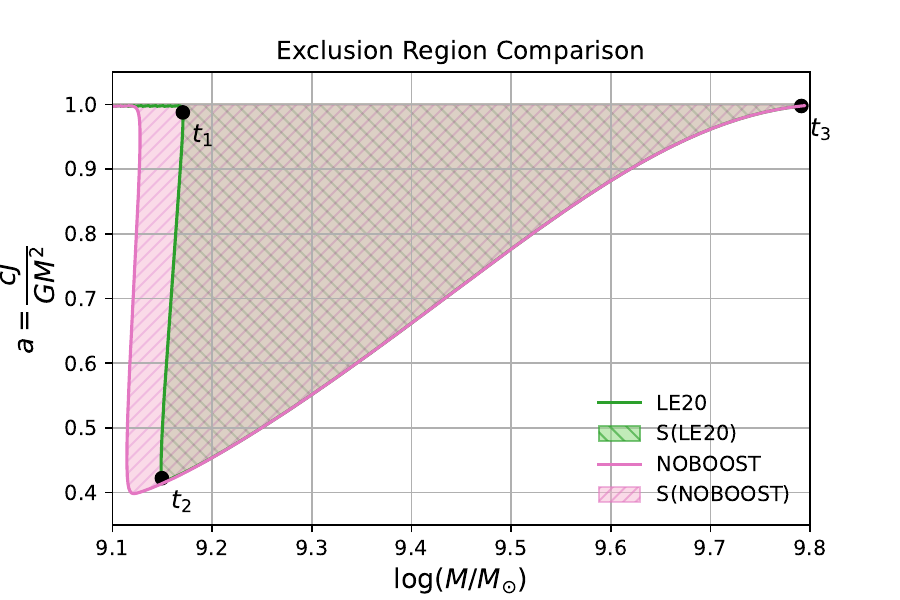}
    \caption{An illustration of the exclusion region. $f_{\rm ex}$ for LE20 is the ratio between the area of the green shaded region and the pink shaded region. $t_1$ is the start of the first superradiance drop, $t_2$ is the end of the first superradiance drop, and $t_3$ is when the black hole is maximally spinning again (the end of the simulation).}
    \label{fig:ex_region_comp}
\end{figure}

The Regge plane exclusion region is illustrated below. The exclusion region of a given simulation is denoted using $S$ with the simulation label as the argument. For example, the exclusion region area of the HM1 case is denoted by S(HM1). The function $S$ is defined as follows:

\begin{multline}
    S = \int_{\log{M(\tau_1)}}^{\log{M(\tau_s)}}a(t=\tau_1,M)d\log M\\
    +(\log M(\tau_2)-\log M(\tau_s))-\int_{\log{M(\tau_1)}}^{\log{M(\tau_2)}}a(t=\tau_2,M)d\log M,
    \label{eq:excl_region_integral}
\end{multline}

where $\tau_s$ and $\tau_1$ are the times for the start of the first superradiance drop and the end of the drop, respectively, as shown in Fig. \ref{fig:ex_region_comp}. $\tau_2$ is the time when $a_{\rm crit} = 1$. As $a$ evolves with time and can have multiple values for a given black hole mass, piecewise functions dependent on the simulation time must be used in the integral.

\subsection{A discussion of relevant timescales}
\label{sec:timescale_intro}

\begin{figure}
\includegraphics[width=\columnwidth]{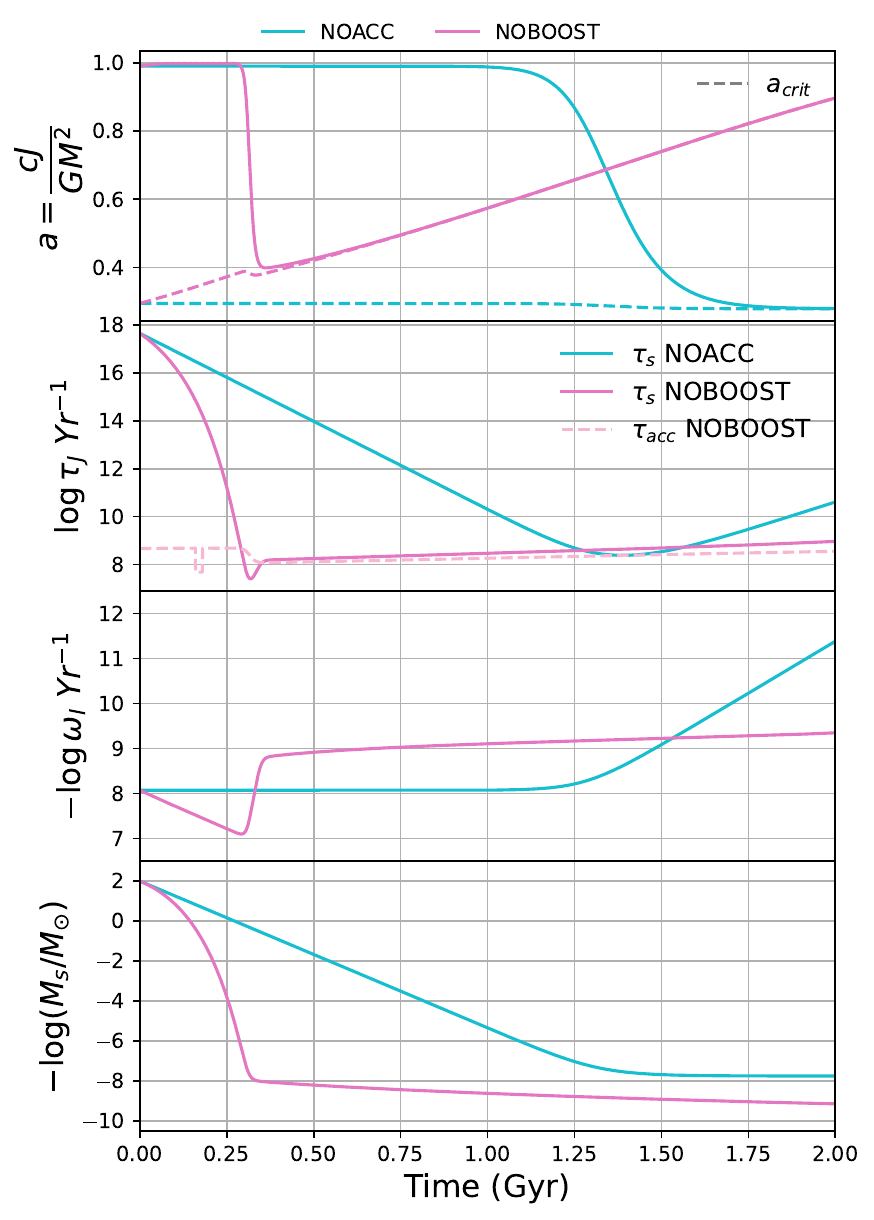}
\caption{Time evolution of, from top to bottom, the characteristic timescales $\tsup$ and $\tacc$, the frequency of the lowest energy orbital $\omega_{\rm I}$, the axion cloud mass $M_{\rm S}$ and the black hole spin $a$ for the case without accretion (blue) and a case of constant accretion of $f_{\rm Edd}-0.05$.}
\label{fig:noboost}
\end{figure}

In this section we explore the impact of superradiance on the spin evolution of black holes under two assumptions: no external mass accretion, called \name{noacc}, and constant mass accretion at an Eddington ratio of $f_{\rm Edd} = 0.05$ called \name{noboost}. As can be seen in Fig. \ref{fig:noboost}, in both cases an initially maximally spinning black hole undergoes a "superradiance drop" where the spin magnitude decreases as angular momentum is extracted from the black hole by the axion cloud. The drop happens earlier with accretion (\name{noboost}) than without (\name{noacc}). To understand why the black hole spin evolves in this manner in both cases it is instructional to consider the two timescales that govern this evolution. 

From Eq. \ref{eq:AMconserveequation}, the timescale for superradiance to exchange angular momentum between the axion cloud and the black hole is 
\begin{equation}
\tsup = \frac{J}{\dot{J}_s}=\frac{Ga}{2c\hbar} M^2 \left ( \frac{M_{\rm S}}{\mu}\right)^{-1} \omega_{\rm I} ^{-1}
\label{eq:tsup}
\end{equation}
where $J=\frac{GM^2a}{c}$. In the absence of accretion, this is the only relevant timescale governing the system. We will discuss \name{noacc} first to understand how the different components in $\tsup$ influence the evolution. 

For constant black hole mass $M$, we can write $\frac{\dot{J}}{J}\propto \frac{M_{\rm S}}{\mu}\omega_{\rm I}$ which means that the frequency term $\omega_{\rm I}$ can also loosely be understood as the rate of angular momentum extraction per axion particle since $\frac{M_{\rm S}}{\mu}$ represents the number of axion particles in the axion cloud. This means the rate of superradiance is higher when there are more axions in the cloud, or when $\omega_{\rm I}$ is larger. 

Taking the log of $\tsup$,
\begin{equation}
    \log{\tsup} = \log(\frac{J}{\dot{J}})=\log(\frac{M^2}{M_{\rm S}})-\log(\omega_{\rm I})+\log(\frac{Ga\mu}{2c\hbar}).
    \label{eq:superradtimescale}
\end{equation}

The first term is a ratio between the mass of the black hole and the scalar cloud whilst the second term is the frequency.

As can be seen in Fig. \ref{fig:noboost}, for our test black hole without accretion (\name{noboost}, blue line) $\tsup$ is originally very high. Initially, $\omega_{\rm I}$ is positive and roughly constant as $a$ changes slowly. As a result, the axion cloud grows in mass (see Eq. \ref{eq:dotMcloud} and third panel in Fig. \ref{fig:noboost}) and $\tsup$ decreases. Once $\tsup$ becomes sufficiently small to be comparable to $\frac{1}{\omega_{\rm I}}$, superradiance efficiently extracts angular momentum from the black hole and $a$ rapidly decreases: We refer to this as the "superradiance drop" throughout the paper. 

At this point there is a turning point in $\tsup$, with the behavior now determined by $\omega_{\rm I}$: As $\omega_{\rm I} \rightarrow 0$, the axion cloud mass stabilizes and $\tsup$ continuously increases. As a result the transfer of angular momentum between the black hole and the axion cloud ceases and the black hole spin stabilised, as $ a \rightarrow a_{\rm crit}$. Since $\log{\omega_{\rm I}}$ does not asymptote to $-\infty$, the black hole spin never exactly reaches the critical spin. It should be noted that $\frac{M^2}{M_{\rm S}}$ at late time adjusts such that it is relatively constant so any shape behaviour in $\tsup$ is driven by $\omega_{\rm I}$. 

The situation can be better understood by considering a discrete picture of fixed time steps. A small $\omega_{\rm I}$ with a black hole spin just above the critical spin means the spin decays slightly closer to the critical spin in a small time step as superradiant extraction still occurs. This then reduces $\omega_{\rm I}$ further through Eq. \ref{eq:freqspin} so that for the next time step, the amount of decay is slightly lower as $\tsup$ increases due to a lower $\omega_{\rm I}$. This process continues and the difference in $a-a_{\rm crit}$ becomes orders of magnitude smaller.

The time at which the drop occurs depends on black hole mass. If either the scalar cloud mass $\rm M_{\rm S}$ or frequency of the black hole $\omega_{\rm I}$ are higher initially then $\tsup$ is lower initially and superradiance effects can be observed faster. Since $\omega_{\rm I}\propto M^8$ and $\tsup \propto \frac{M^2}{\omega_{\rm I}}=M^{-6}$ from Eq. \ref{eq:tsup}, increasing black hole mass greatly reduces $\tsup$ and therefore the characteristic timescale on which the problem evolves. By contrast, $a$ changes on the order of unity so the initial spin (and the evolution of the spin throughout) has little influence on the evolution of the system. $M$ and $\rm M_{\rm S}$ remain the dominant variables.

In the presence of accretion, the timescale of accretion also becomes relevant to the evolution of the system: 
\begin{equation}
\tacc = \frac{J}{\dot{J}_{\rm acc}} 
\label{eq:tacc}
\end{equation}
is the timescale for the black hole to obtain angular momentum through external mass accretion, where $\dot{J}_{\rm acc}$ follows Eq. \ref{eq:Jaccterm}.

Now consider external baryonic Eddington accretion onto the black hole in this timescale framework. As can be seen in Fig. \ref{fig:noboost}, in the presence of accretion (\name{noboost}) the super-radiance drop occurs earlier and the black hole spin $a$ is not approximately constant after the drop. Notably, the superradiance drop occurs in the brief window while $\tsup<\tacc$. From Eq. \ref{eq:AMconserveequation}, once $\dot{J}_{\rm ACC} < \frac{\hbar \dot{M}_{\rm S}}{\mu}$ the black hole starts losing angular momentum and $a$ decreases, at first slowly and then increasingly rapidly as $\tsup \rightarrow \frac{1}{\omega_{\rm I}}$ when again a superradiance drop occurs. The sped-up evolution is mostly driven by an increased rate of growth for $M_{\rm S}$ which causes $\tsup$ to decrease faster. There is also a smaller increase in $\omega_{\rm I}$ (decrease in $-\log{\omega_{\rm I}}$ before the superradiance drop, not present in \name{noacc} which also speeds up the evolution of the system.

Unlike in \name{noacc}, the spin $a$ in \name{noboost} is not constant after the drop but increases again with time. However, phenomenologically, the two systems behave similarly, as this increasing spin at late times is driven by the fact that $a_{\rm crit}(M)$ and the critical spin continues to increase as the black hole grows. During the superradiance drop, $\omega_{\rm I}$ rapidly decreases ($-\log{\omega_{\rm I}}$ increases) but then, unlike in \name{noacc}, levels off longterm. This reflects the fact that a state of equilibrium is established between the growth of the axion cloud and that of the black hole. During each accretion event, the black hole finds itself at $a > a_{\rm crit}(M)$ i.e. the accreted angular momentum spins up the black hole above the critical spin for its current mass. The change in $a_{\rm crit}$ also explains why the drop in $a$ is smaller for \name{noboost} than for \name{noacc}: by the time the superradiance drop occurs, the black hole mass M has grown in \name{noboost}, and as a result $a_{\rm crit}$ is higher. As $\tsup \gtrsim \tacc$, superradiance  extracts that extra angular momentum, driving the longterm (albeit slow) evolution in $M_{\rm S}$ that can be seen in Fig. \ref{fig:noboost}, and keeping $a \sim a_{\rm crit}$ over long periods of time. 

Differentiating Eq. \ref{eq:acrit} gives the rate of increase in critical spin for a given accretion rate as:

\begin{equation}
    \frac{da_{\rm crit}}{dt}=\frac{4(1-4\alpha^2)}{(1+4\alpha^2)^2}\frac{G\mu}{c\hbar}\frac{dM}{dt}.
    \label{eq:derivacrit}
\end{equation}

$dM/dt\approx \dot{M}_{\rm ACC}$ during the late time evolution since the mass extraction rate due to the effect superradiance is at a maximum $\sim\frac{1}{100}$ of the mass accretion rate. 

As the black hole mass continues to increases, $\tsup$ and $\tacc$ continue to balance each other closely as the black hole slowly spins up, follow the critical spin trajectory. The plateauing of $\tsup$ after the superradiance drop means the frequency term $\omega_{\rm I}$, which defines the evolution of the $\tsup$ at late times, is relatively constant (we call this the equilibrium frequency). If the black hole spin were to deviate more significantly from $a_{\rm crit}$ then $\omega_{\rm I}$ term would correspondingly increase. The equilibrium value of $\omega_{\rm I}$ found here is likely set by a combination of our choice of $f_{\rm Edd}$ and the timestep we chose for integration, as this determines the total angular momentum injected into the black hole at each accretion event. 



If the spin of the black hole were to evolve such that $a_{\rm crit} > a$ at any time, then $\omega_{\rm I}$ would become negative. Physically, in this case the axion cloud (and its momentum) would accrete back onto the black hole, spinning it back up until $a \sim a_{\rm crit}$. Again, the effect becomes less efficient as $a \rightarrow a_{\rm crit}$ so the black hole spin only ever asymptotes towards 


A note to mention is the use of an initial seed mass for the axion cloud. For our simulations we have used a seed mass of $0.01 \rm M_{\odot}$ throughout. A seed mass of $0.01 \mu$ could also be used to mimic a quantum fluctuation in the axion cloud. Since the superradiance drop occurs when $M_{\rm S}$ is large enough, the seed mass has a direct impact on the exclusion region. Generally a lower axion mass means $\tsup$ begins at a higher value meaning it takes longer to decay to $\sim \frac{1}{\omega_{\rm I}}$ which in turn allows more mass to be accreted into the black hole reducing the exclusion region. For reference (not shown) in the constant accretion model of $f_{\rm Edd} = 0.05$, a seed mass of $0.01 \mu$ causes the superradiance drop to occur at $\sim 10^{9.2} M_{\odot}$ compared to $10^{9.1}M_{\odot}$ for a seed mass of $0.01 M_{\odot}$.

\section{Results}

In this section, we test the impact of the timing of a single accretion boost ($t_{\rm ev}$) and the length of the accretion boost ($\Delta t_{\rm ev}$) on the evolution of the black hole in the Regge plane. Unless otherwise specified, the axion mass is  $10^{-20}eV$. Black holes initially grow at $f_{\rm Edd} = 0.05$.  Accretion is boosted at time $t_{\rm ev}$ for a duration of $\Delta t_{\rm ev}$ and then returns to $f_{\rm Edd} = 0.05$. 

Boosted simulations are compared to the simulation \name{noboost} from Section \ref{sec:timescale_intro}  that retains $f_{\rm Edd} = 0.05$ throughout.

\subsection{Low boost}
\label{sec:weak_boost}


\begin{figure*}
	\includegraphics[width=\columnwidth]{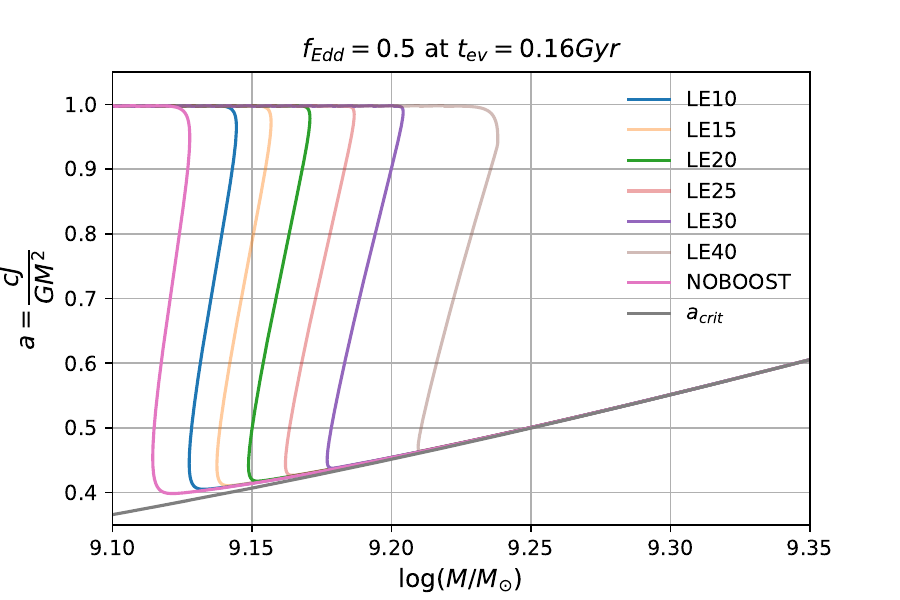}
    \includegraphics[width=\columnwidth]{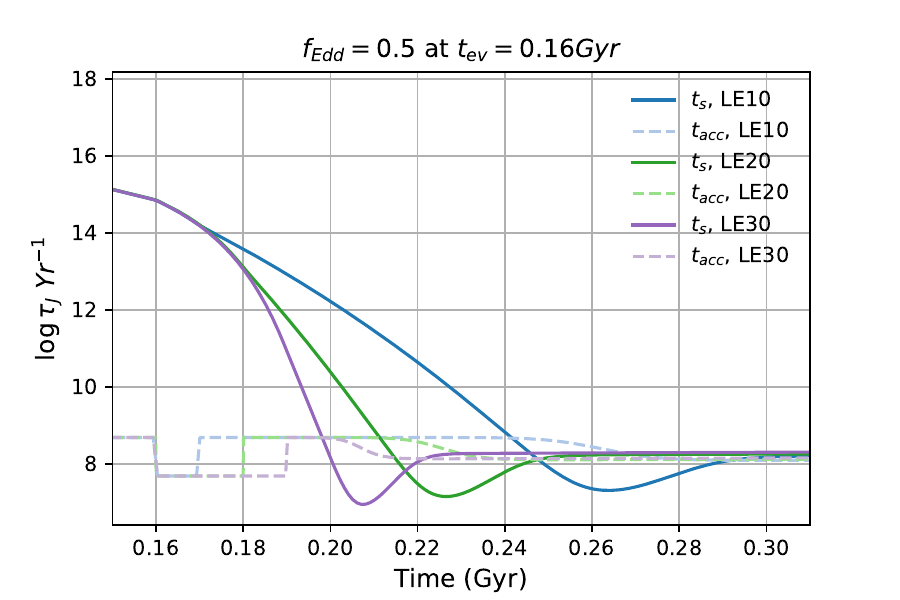}
    \includegraphics[width=\columnwidth]{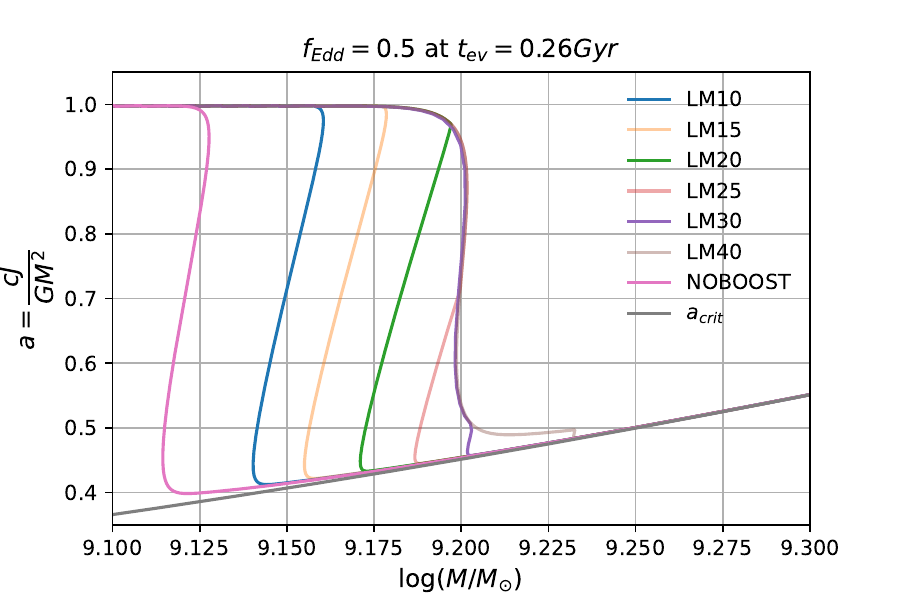}
    \includegraphics[width=\columnwidth]{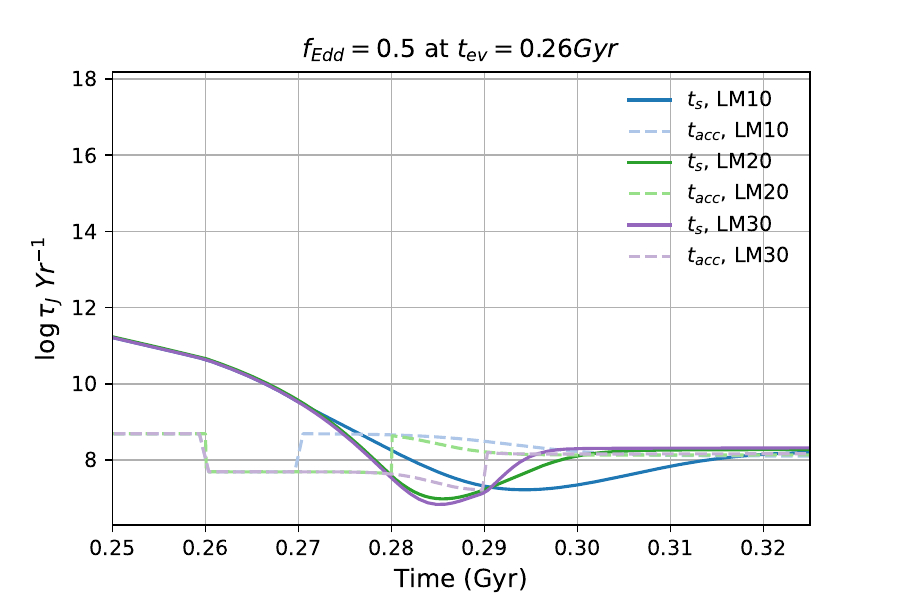}
    \includegraphics[width=\columnwidth]{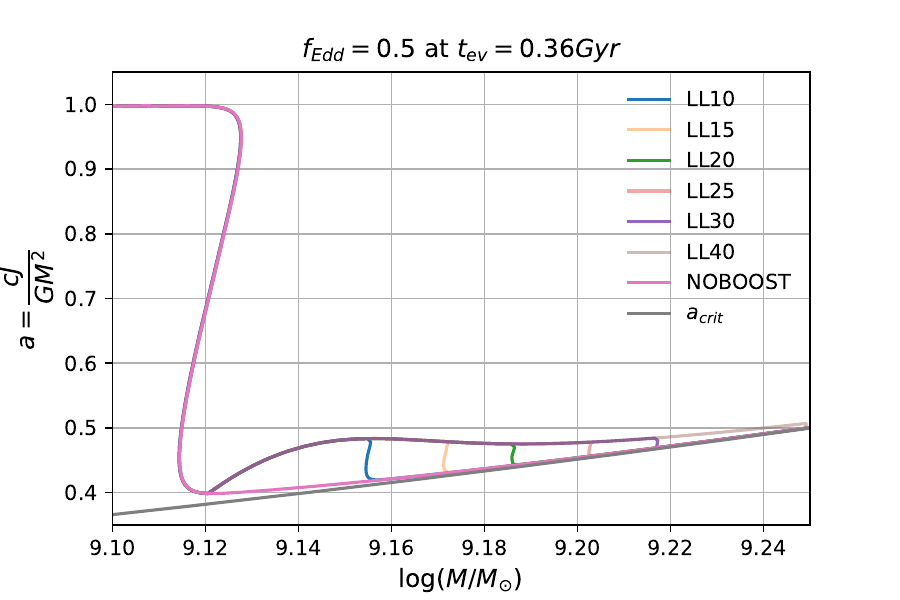}
    \includegraphics[width=\columnwidth]{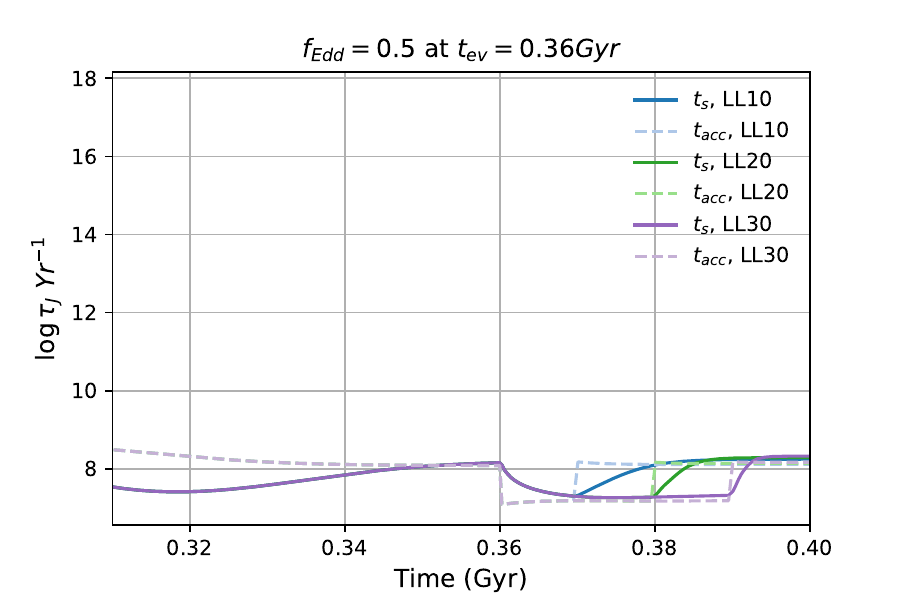}
    \caption{{\it Left:} The Regge plane shows that as $\Delta t_{\rm ev}$ increases, more of the exclusion region closes off. For boosts applied before the initial superradiance drop, the drop itself is pushed to higher mass levels the longer the boost. This will tend to the constant accretion trajectory for $f_{\rm Edd}=0.5$. For boosts applied after the initial superradiance drop ($t_{\rm ev} = 0.36Gyr$) the boost deviates the black hole away from the critical spin trajectory. A second drop in spin occurs when the boost is released, so the longer the duration, the longer the black hole remains deviated. {\it Right:} A longer duration boost applied before the superradiance drop means more mass is accreted into the black hole causing quicker decay in timescale as $\omega_{\rm I}$ is larger at a given time. Once the drop occurs, $\tsup$ establishes an equilibrium with the external mass accretion. For boosts applied after the superradiance drop, $\tsup$ is not faster than $\tacc$ until the boost is removed. With a larger $\Delta t_{\rm ev}$ in this regime, the black hole continuously tends to the critical spin trajectory without ever losing angular momentum, establishing a new equilibrium with the $f_{\rm Edd} = 0.5$ accretion rate.}
    \label{fig:fedd05_timing}
\end{figure*}

First, we analyse the impact of a low (\name{L}) boost to $f_{\rm  Edd} = 0.5$, a factor 10 in comparison to the baseline accretion. In this section we analyse three timings for the accretion boost: an early timing boost at $t_{\rm ev} = 0.16 \rm \ Gyr$, labeled \name{LE}, an intermediate boost at $t_{\rm ev} = 0.26 \rm \ Gyr$ labeled \name{LM}, and a late boost that occurs at $t_{\rm ev} = 0.36 \rm \ Gyr$ labeled \name{LL} following the naming convention established in Sec. \ref{sec:params_names}. Without the accretion boost, the super-radiance drop occurs at $t \approx 0.29 $ Gyr.

As can be seen in Fig. \ref{fig:fedd05_timing}, even short accretion boosts that start and end before the superradiance drop
reduce the exclusion region. The timing of the onset and the duration of the accretion boost significantly affect the shape of the exclusion region in the Regge plane. In comparison to the \name{noboost} case, all accretion boosts studied here reduce the size of the exclusion 
regions but the timing of the onset of the boost ($t_{\rm ev}$) critically matters. The largest reduction is seen for the \name{lM} simulations (middle panels), whose boost begins just before the superradiance drop in the \name{noboost} case. Any impact of boosts is temporary, as all simulations evolve along the critical spin at late times. Generally across all $t_{\rm ev}$, longer accretion boosts (larger $\Delta t_{\rm ev}$) close off more of the exclusion region. For boosts applied before the superradiance drop (simulation sets \name{E} \& \name{M}), a larger $\Delta t_{\rm ev}$ means more mass is accreted into the black hole before the drop. 

The right-hand panels of Fig. \ref{fig:fedd05_timing} show the time evolution of $\tsup$ and $\tacc$ for a selected subset of simulations. The time of the boost can be identified by the temporary discontinous drop in $\tacc$. The late continuous decrease in $\tacc$ is due to the spindown of the black hole following the superradiance drop. As discussed in Sec. \ref{sec:timescale_intro}, the superradiance drop occurs when $\tacc \sim \tsup$.

As can be seen in the right-hand panel of Fig. \ref{fig:fedd05_timing}, during the boost the decay rate of $\tsup$ increases because $\omega_{\rm I}$, which is strongly dependent on the black hole mass, grows faster than without the boost. This means $\omega_{\rm I}$ is higher during the boost at a given time than without the boost. A higher $\omega_{\rm I}$ causes $M_{\rm S}$ to grow faster, which in turn leads to a steeper reduction in $\tsup$ even after the boost has finished. As mentioned in Sec \ref{sec:timescale_intro}, $\omega_{\rm I}$ is relatively constant during this early period compared to $M_{\rm S}$, which means the moment the superradiance drop occurs is when $M_{\rm S}$ reaches a critical mass, driven by the external accretion. This means that the superradiance drop (where $\tsup \sim \tacc$) occurs earlier in time for longer accretion bursts, but at a higher black hole mass (see left-hand panel of Fig. \ref{fig:fedd05_timing}). The higher mass of the black hole at the superradiance drop means the black hole meets the critical spin evolution track at a higher value of spin. Longer boosts close off more of the accretion region as they maintain the black hole at a maximum spin state for longer.

Comparing the first and second row of Fig. \ref{fig:fedd05_timing}, it can be seen that equivalent boosts applied at $t_{\rm ev} = 0.26 \ \rm Gyr$ (\name{M}) compared to $t_{\rm ev} = 0.16 \ \rm Gyr$ (\name{E}) closes more of the exclusion region.  Hence boosts applied closer to the original superradiance drop have a larger impact on the exclusion region than those further away. 
The boosts bring the black hole mass to the point where the superradiance drop occurs (in the mass domain). There is a time delay between the onset of the superradiance drop and the end of the boost but not much mass is added in this phase for a noticeable difference in the Regge plane. Therefore, the larger the mass of black hole at $t_{\rm ev}$ the larger the black hole mass will be at the end of the boost for a given $\Delta t_{\rm ev}$. This means a larger $t_{\rm ev}$ will lead to a larger shift in the Regge plane. $t_{\rm ev} + \Delta t_{\rm ev}$ should be less than the time of the second superradiance drop for the above logic to hold. In cases where it is shorter, the boost saturates on the drop (as seen in \name{LM20}, \name{LM25}, \name{LM30}, and \name{LM40}).

For some of the $t_{\rm ev} = 0.26 \rm \ Gyr$ boosts ( \name{M}, middle panels of Fig. \ref{fig:fedd05_timing}), the boost is still active when the superradiance drop occurs (namely \name{LM20}, \name{LM25} and \name{LM30}). These simulations converge onto where the drop would have been for a simulation with constant $f_{\rm Edd} = 0.5$. As soon as the boost ends (see e.g. \name{LM20} just after the onset of the drop) they evolve away again as the accretion timescale increases again until a new equilibrium is restored when the black hole mass and spin has evolved back onto the critical spin trajectory. This shows that for a black hole with variable accretion, it is the peak accretion rates that determine the size and shape of the remaining exclusion region. Unlike for the early boosts in the top panel, boosts that occur during the superradiance drop are self-limiting in the sense that there is a maximum reduction in the exclusion region for any boost length studied here.


 
Now consider late time accretion events applied after the initial superradiance drop at $t_{\rm ev} = 0.36Gyr$ (\name{L}, bottom row of Fig. \ref{fig:fedd05_timing}). In this regime $\tacc$ and $\tsup$ are at an equilibrium point prior to the boost, as explained in section \ref{sec:timescale_intro}. Boosting the accretion in this regime disrupts the established equilibrium as the accretion timescale $\tacc$ drops during the accretion boost (see bottom right panel of Fig. \ref{fig:fedd05_timing}). This evolves the black holes away from the critical spin temporarily. During this time, $\tsup$ decreases until it establishes a new equilibrium state with the new accretion rate. Therefore a greater difference in $\tacc$ and $\tsup$ (by a greater accretion boost) would require $\tsup$ to decay further. This gives more time for accretion to spin up the black hole, deviating it further away from the critical spin trajectory. 

When the spin deviates away from the critical spin, $\omega_{\rm I}$ is the only term directly affected since $\omega_{\rm I}\propto (a-a_{\rm crit})$. Therefore the response to bring the system back to equilibrium is driven by $\omega_{\rm I}$. A larger $\omega_{\rm I}$ would lead to a stronger response from the black hole to bring $\tsup$ to the new equilibrium state. From considering the Eq. \ref{eq:freqspin} form of $\omega_{\rm I}$, a fixed difference $a-a_{\rm crit}$ with a higher mass black hole would result in a higher $\omega_{\rm I}$ causing a stronger 'restoring response'. Therefore higher-mass black holes do not deviate as much from $a_{\rm crit}$ for a fixed boost within this regime.

In the above results, only the \name{LL40} boost lasts long enough for the new equilibrium to be established. When the shorter boosts are released, a second small super-radiance drop occurs as $\tsup$ dominates over $\tacc$, which re-establishes equilibrium with the initial accretion rate.

Overall, for the set of simulations studied in this section we see the largest reduction in the accretion region for the longest boosts that occur before the superradiance drop actually occurs. Once the drop begins, the shape of the exclusion region is bounded by that of constant accretion at the boosted accretion rate. Boosts after the superradiance drop can temporarily reduce the she exclusion region by increasing the black hole spin. However, for all boosts tested here the black hole spin remains moderate if the boost occurs after the accretion boost.


\subsection{Stronger boosts}
\label{sec:strong_boost}

We now test the impact of boosts of a factor 100 over the baseline accretion rate, to $f_{\rm Edd} = 5$ (simulation suite \name{H}). This represents a boost that is 10 times stronger than those discussed in Sec. \ref{sec:weak_boost} (simulation suite \name{L}). We have reduced the boost lengths by a factor 10 in comparison to those probed in Sec. \ref{sec:weak_boost} to ensure that the total mass accreted during boosts remains the same. This choice also reflects the fact that simulations have found that higher Eddington accretion rates can generally be sustained for shorter periods of time (see Sec. \ref{sec:introduction} for a discussion)

\begin{figure*}
	\includegraphics[width=\columnwidth]{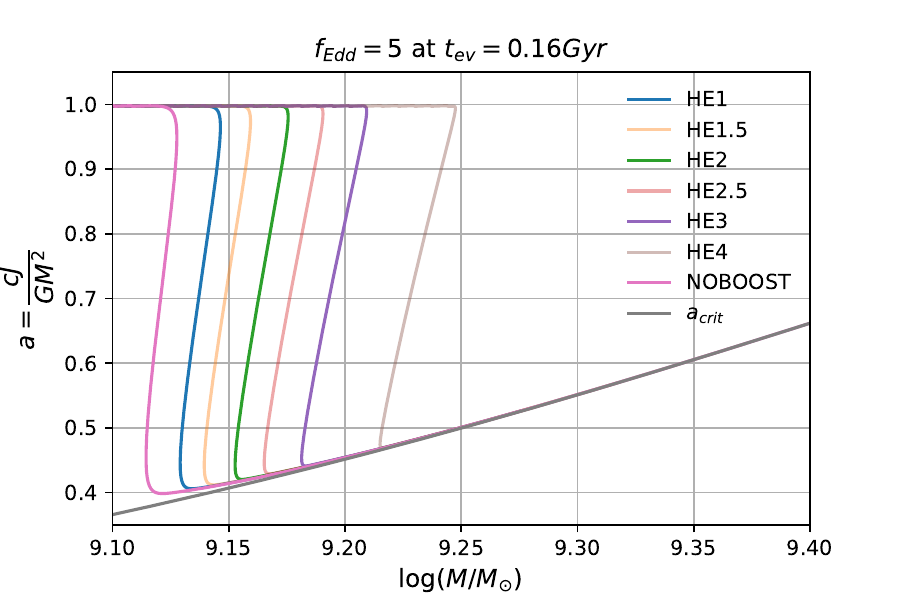}
    \includegraphics[width=\columnwidth]{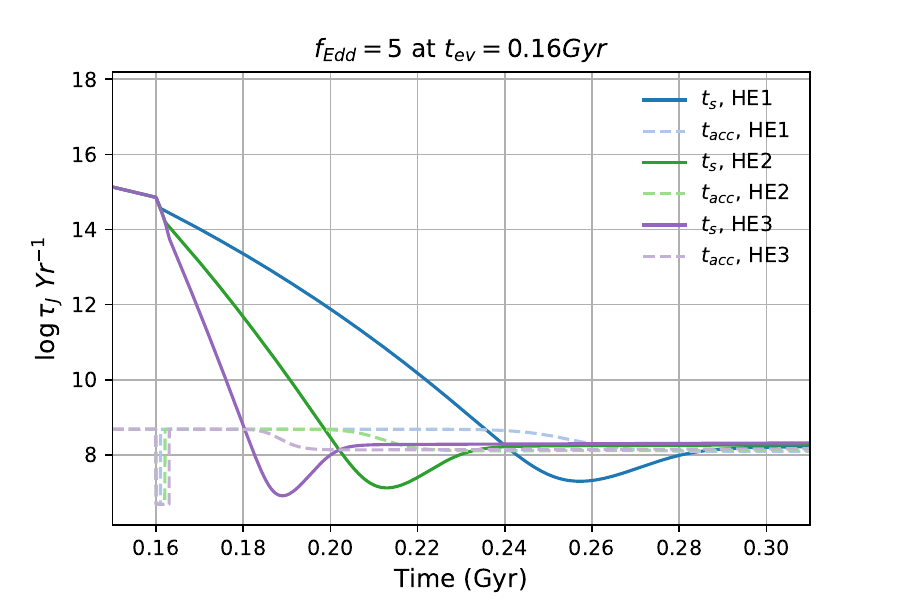}
    \includegraphics[width=\columnwidth]{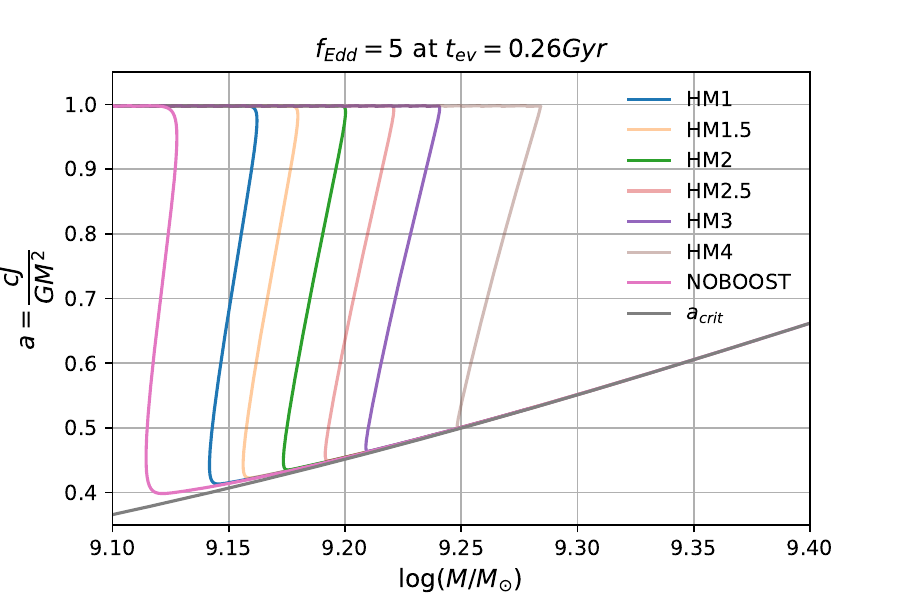}
    \includegraphics[width=\columnwidth]{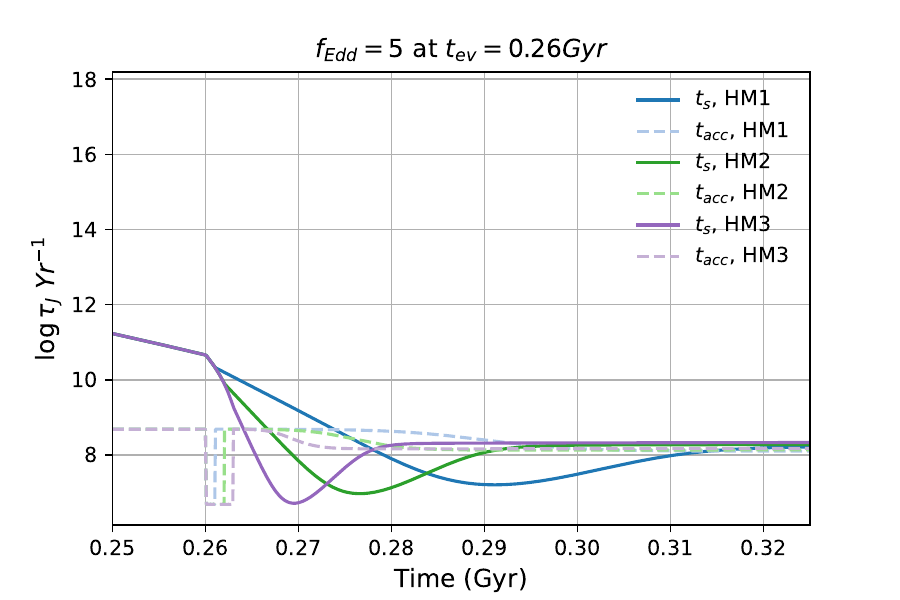}
    \includegraphics[width=\columnwidth]{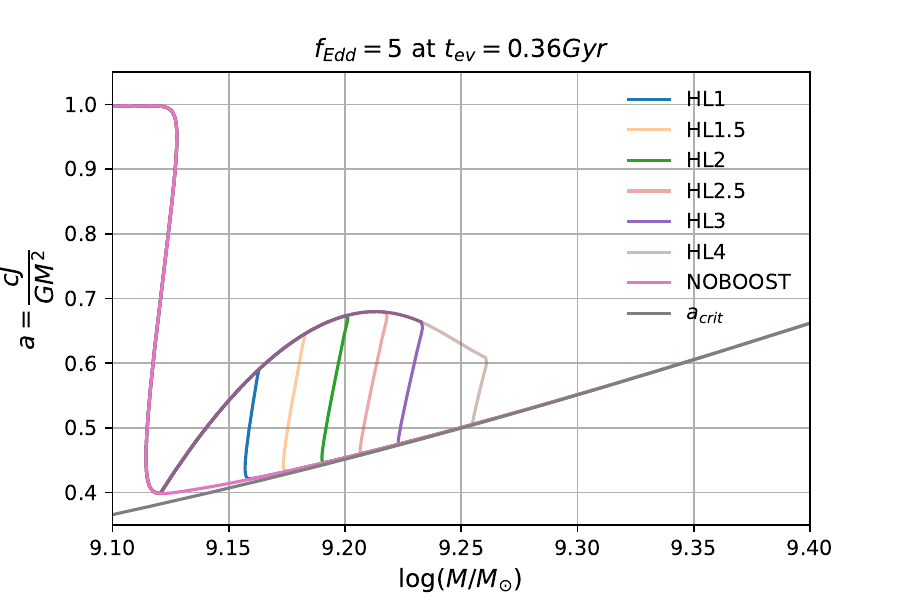}
    \includegraphics[width=\columnwidth]{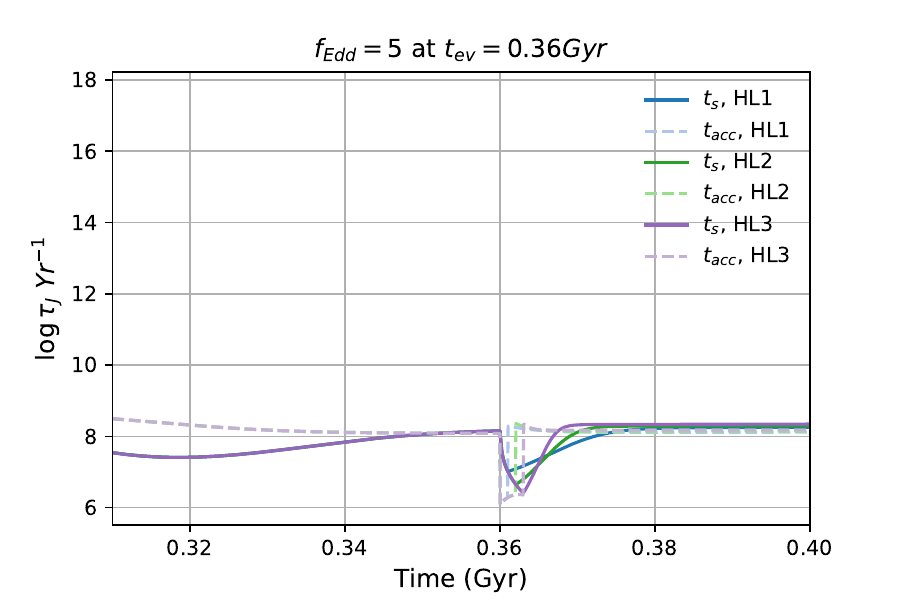}
    \caption{{\it Left:} The Regge trajectory can be seen to be similar to the $f_{\rm Edd} = 0.5$ model.{\it Right:} The timescale decay can be seen to occur much faster than in the $f_{\rm Edd} = 0.5$ boost case since the rate at which the black hole grows is faster, decreasing $t_s$ quicker. }
    \label{fig:fedd5timing}
\end{figure*}

In comparison to those shown in Fig. \ref{fig:fedd05_timing}, the Regge trajectories for stronger boosts applied at the same $t_{\rm ev}$  in Fig. \ref{fig:fedd5timing} before the initial superradiance drop (\name{HE} and \name{HM} versus \name{LE} and \name{LM}) share a similar shape. This reflects the fact that both sets of black holes are at a similar mass at the onset of the superradiance drop.

To understand the differences, we must consider the value of $M_{\rm S}$ at the end of the boost, since $M_{\rm S}$ drives the evolution to superradiance. Applying chain rule on Eq. \ref{eq:dotMcloud} and assuming $\frac{dM}{dt}\approx \dot{M}_{\rm acc}$ we get:
\begin{equation}
    \frac{dM_{\rm S}}{dM} = \frac{2M_{\rm S}\omega_{\rm I}}{\dot{M}_{\rm acc}} = \frac{2M_{\rm S}\omega_{\rm I}}{f_{\rm Edd}M}
\end{equation}

Integrating from a black hole mass of $M_0$ to $M_1$ gives :
\begin{equation}
    M_{\rm S}(M_1)=M_{\rm S}(M_0)\exp\left({\frac{2}{f_{\rm Edd}}\int_{M_0}^{M_{\rm S}(M_1)}\frac{\omega_{\rm I}(M,a)}{M} dM}\right)
    \label{eq:axcloud_variation}
\end{equation}
All terms in the integral only depend on $M$ since $a$ can be considered to be constant before the superradiance drop. The integral will therefore be the same at the end of the applied boost for both strong and weak boosts (since the total mass accreted is the same). Therefore the $\frac{1}{f_{\rm Edd}}$ factor reduces the axion cloud mass at the end of the boost.
So, for the lower boost (boost to $f_{\rm Edd} = 0.5$, Fig. \ref{fig:fedd05_timing}) the axion cloud is much larger at the end of the boost than for the stronger boost (boost to $f_{\rm Edd} = 5$, Fig. \ref{fig:fedd5timing}) since it has had more time to grow. Due to the smaller axion cloud at the end of the boost it takes longer from the end of the boost to when superradiance dominates. This additional time between the end of the boost and the superradiance drop means more mass is accreted at the baseline accretion rate before the superradiance drop for the strong boosts than the weak boosts so the drop in spin occurs at a somewhat higher black hole masses. This difference in mass is comparatively small, as the black hole is heavier than before the boost so axion cloud mass increases quickly (due to higher $\omega_{\rm I}$), driving $\tsup$ down, while the accretion rate is again low (the baseline accretion rate $f_{\rm Edd} = 0.05$). These effects mean there is not enough time for the mass of the black hole to significantly increase compared to the equivalent $f_{\rm Edd} = 0.5$ boost case, causing both trajectories to be similar in the Regge plane.

In the middle panels of Fig. \ref{fig:fedd5timing} the short boost durations for all simulations tested here mean that all boosts finish before the onset of the superradiance drop. For this reason the \name{HM} simulations show no evidence of the convergent behaviour seen for e.g. \name{LM25}. 

For boosts applied after the initial superradiance drop at $t_{\rm ev} = 0.36 \rm \ Gyr$ (\name{HL}), we can see higher $f_{\rm Edd}$ boosts lead to larger deviations from $a_{\rm crit}$  than for the \name{LL} simulations (compare bottom left panels of Fig. \ref{fig:fedd5timing} and \ref{fig:fedd05_timing}). As explained in Sec. \ref{sec:timescale_intro} the deviation occurs because the equilibrium is disrupted and $\tsup$ must decay further to re-establish the equilibrium (see also left hand panel of Fig. \ref{fig:fedd5timing}). A higher $f_{\rm Edd}$ boost means the initial difference between $\tacc$ and $\tsup$ is larger. Therefore a higher $f_{\rm Edd}$ boost at a specific $t_{\rm ev}$ after the initial superradiance drop will cause the spin of the black hole to deviate further. With a larger deviation, $\omega_{\rm I}$ increases, decreasing $\tsup$ directly through Eq. \ref{eq:tsup}. However, this effect is insignificant since the difference between the $a-a_{\rm crit}$ term in the two cases is smaller than an order of magnitude for a given mass. As in the case in Sec. \ref{sec:weak_boost}, $M_{\rm S}$ growth (through $\omega_{\rm I}$) drives the reduction of $\tsup$. The spin deviates further from $a_{\rm crit}$ until either $M_{\rm S}$ grows large enough to reduce $t_s$ and establish a new equilibrium, or the boost is removed in which $\tacc$ will jump to a higher value causing a second superradiance drop. While the boost is still active, the Regge trajectory can be seen to fall back to $a_{\rm crit}$ much more gradually (the process of equilibrating with the boosted accretion rate) than the initial superradiance drop. This is because $\omega_{\rm I}$ is much lower since $a-a_{\rm crit}$ is lower compared to before the initial superradiance drop. This prevents an overshoot in $M_{\rm S}$ growth as seen in the initial superradiance effect since $\omega_{\rm I}$ tends to more modest values as equilibrium is re-established. 

Overall we conclude that for boosts before the superradiance drop it is predominantly the total mass accreted that determines the evolution of the black hole in the Regge plane and as a result the effective exclusion region. The evolution of black hole constantly accreting at the peak accretion provides a limiting case that determines the absolute limits of the accretion region. All black holes temporary boosted to that accretion rate predict an accretion rate larger than for peak constant accretion. 

For accretion boosts after the superradiance drop, even strong sustained accretion boosts only produce a comparatively moderate deviation from the critical spin. The magnitude of the maximum deviation is set by the peak accretion rate, with larger Eddington ratios producing larger deviations. Longer boosts do not increase the maximum spin of the deviation but they do decrease the size of the exclusion region further as they maintain black holes at increased spin for longer. Once the boost stops, black holes drop back onto the critical spin at almost constant black hole mass in a second, smaller superradiance drop.

\subsection{Varying axion mass}
\label{sec:varying_ax_mass}

\begin{figure*}
	\includegraphics[width=\columnwidth]{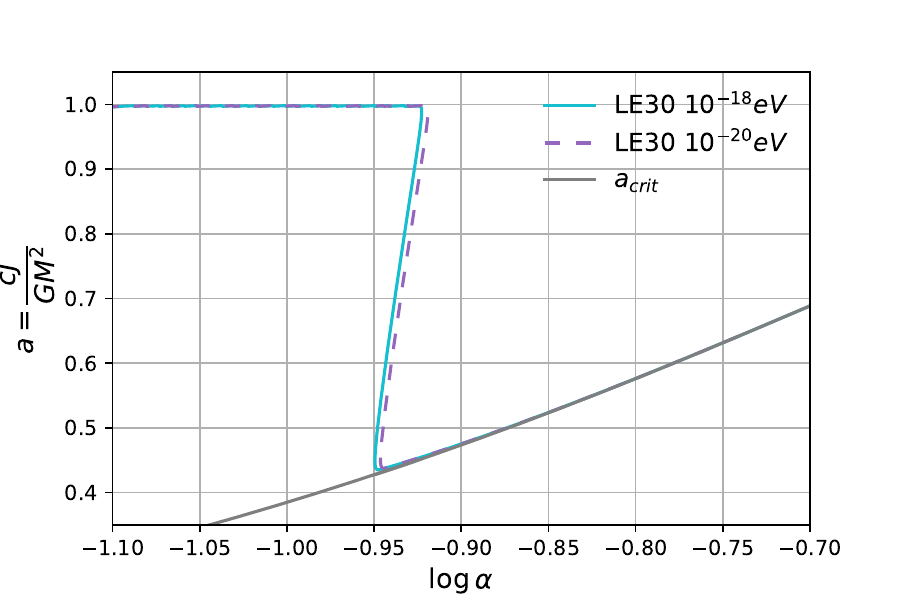}
    \includegraphics[width=\columnwidth]{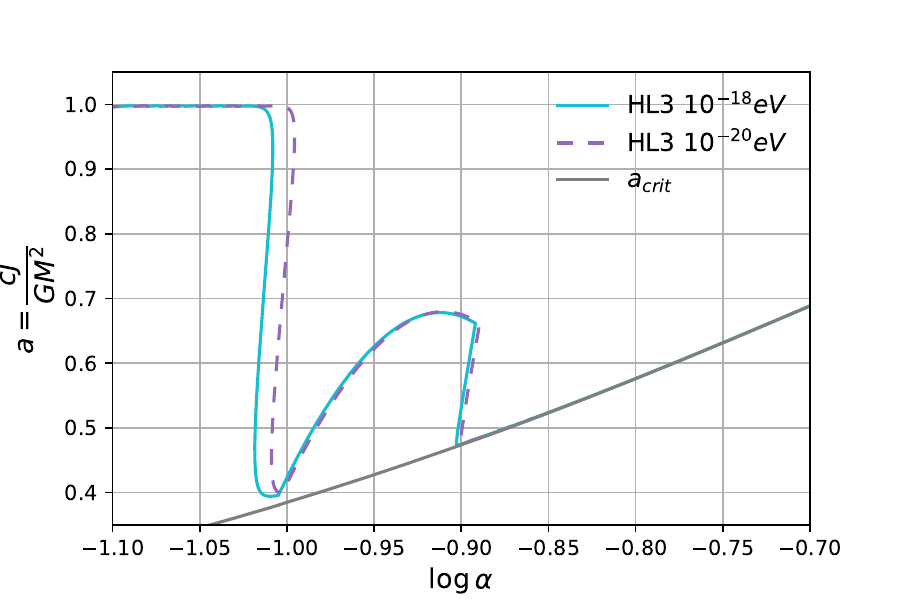}
    \caption{Regge trajectory for simulations with axion mass of $10^{-18}eV$ (cyan, \name{highaxion}) and $10^{-20}eV$ (purple dashed simulation, \name{LE30}) respectively.  The parameters of \name{highaxion} are set so that all $f_{\rm Edd}$ ratios are increased by a factor of 100 and any time value and duration is reduced by a factor of 100 (which is equivalent to the ratio $10^{-18}eV$ and $10^{-20}eV$) compared to the fiducial simulation  \name{LE30}. This means that \name{highaxion} has a background accretion rate of $f_{\rm Edd} = 5$, which is boosted to $f_{\rm Edd} = 50$ at $t_{\rm ev} = 0.16Myr$ for $\Delta t_{\rm ev} = 30Kyr$. The overlap of the trajectories shows the universality of superradiance, and how changing the accretion parameters is degenerate with changing the axion mass. }
    \label{fig:high_axion_mass}
\end{figure*}

So far we have only analysed simulations with an axion mass of the $10^{-20}eV$. In this Section we will use a variant of the simulation \name{LE30} with an axion mass of $10^{-18}eV$, which we will call \name{highaxion}. The Regge trajectories were found to coincide when the parameters of \name{highaxion} are set so that all ratios $f_{\rm Edd}$ are increased by a factor of 100 and any time value and duration are reduced by a factor of 100 (which is equivalent to the mass ratio between axion masses of $10^{-18}eV$ and $10^{-20}eV$) compared to the fiducial simulation \name{LE30}. This means that \name{highaxion} has a background accretion rate of $f_{\rm Edd} = 5$, which is boosted to $f_{\rm Edd} = 50$ at $t_{\rm ev} = 0.16 \rm \ Myr$ for $\Delta t_{\rm ev} = 30 \rm  \ Kyr$. Increasing the accretion efficiency but decreasing the duration means that the total mass accreted onto both black holes remains the same.  Fig. \ref{fig:high_axion_mass} shows how adjusting the accretion ratios and durations in this way compensates for the change in axion mass. This shows a universality in the superradiance effect where a change in accretion efficiency can compensate for a change in axion cloud mass as long as the total accreted mass remains the same. 

 To understand why a higher axion mass needs more efficient accretion to produce the same reduction of the exclusion region, we investigate how the timescale for the onset of accretion is affected by the change in axion mass $\mu$. The derivation using a constant accretion model and natural units ($c=1,G=1,\hbar=1$) is outlined in detail in App. \ref{sec:appendix_superradtime}. Denoting $t_s$ as the time to superradiance, the final result shows that for different axion masses of $\mu_1$ and $\mu_2$ we expect 
\begin{equation}
    t_{s,1} = \frac{A_2}{A_1}\left[t_{s,2} -\frac{24}{A_2}\log\left(\frac{\alpha_1^6M_{1,S}\mu_1^2f_{2,\rm Edd}(a_1-a_{1,\rm crit})}{\alpha_2^6M_{2,S}\mu_2^2f_{1,\rm Edd}(a_2-a_{2,\rm crit})}\right)\right]
    \label{eq:axion_mass_delay}
\end{equation}
where $A_i=(a_i-a_{i,\rm crit})\alpha_i^8\mu_i$, $a_i$ and $a_{i,\rm crit}$ are dimensionless spin and critical spin parameters , $M_{i,S}$ are the initial cloud masses, and $f_{i,\rm Edd}$ are the Eddington ratios of the accretion associated with two axion masses of $\mu_i$, where $i=1,2$. All terms in Eq. \ref{eq:axion_mass_delay} are evaluated at $t=0$.
Note that in deriving Eq. \ref{eq:axion_mass_delay} we assume the black hole to be maximally spinning due to $\tacc<<\tsup$ initially. This means the black hole will always spin up maximally long before the superradiance drop. Hence, $a_1\approx a_2$ is usually the case, but we've kept the terms in the equation for completeness.
This indicates that altering the axion mass affects the onset of superradiance in two ways simultaneously. First there is a simple scaling (the ratio of $A_2/A_1$ on Eq. \ref{eq:axion_mass_delay}) in which higher mass axions experience the superradiance drop early by a factor of $\mu_2/\mu_1$, given a constant initial $\alpha$ in both cases. Secondly, the final term imposes a further time delay in the onset of the superradiance drop. When $f_{2,\rm Edd} = \frac{\mu_2}{\mu_1}f_{1,\rm Edd}$ and initial $\alpha_1=\alpha_2$ as in our simulation scheme, the final term is generally much smaller than $t_{s,2}$. If black holes both start at the same spin with matching initial $\alpha$ and axion cloud seed mass, then we approximately get $t_{s,1}\approx \frac{\mu_2}{\mu_1}t_{s,2}$. This scales all characteristic timescales by the axion mass ratios. Since mass accretion is largely unaffected by superradiance effects before the superradiance drop, $\alpha_s = \alpha_0 \exp{(f_{\rm Edd}t_s)}$ at the point of the superradiant drop which means the $\alpha_s$ at which the superradiance drop occurs is the same for both axion masses given the above conditions. An accretion boost can be considered as another regime of constant accretion, meaning the same time scalings hold as seen in Fig. \ref{fig:high_axion_mass}. Further details of this are explained in App. \ref{sec:appendix_superradtime}.

One point that should be noted is the wider difference in the initial \name{noboost} superradiant drop of the hL3 regge trajectory compared to the lE30 case. This is due to the negative time delay term in Eq. \ref{eq:axion_mass_delay} as already mentioned above. The delay leads to a slightly earlier $t_{s,1}$ (assuming $\mu_1>\mu_2$), which leads to a lower $\alpha$ at which the superradiance drop occurs relative to the $\mu_2$ case. The difference is more pronounced in the \name{noboost} case since $\alpha$ is lower at the superradiance drop, leading to a larger time delay in Eq. \ref{eq:axion_mass_delay}.

It can be seen in the second panel of Fig. \ref{fig:high_axion_mass} that after the initial superradiance drop, in the late spin evolution regime, the spins of the BH's align to the critical spin trajectory. This the case where an accretion boost is applied after the initial superradiance drop.  Therefore, to prove the alignment of the trajectories in the Regge space, we must show that $\frac{da}{d\ln\alpha}$ is independent of $\mu$. The mathematical derivation of this is laid out in App. \ref{sec:appendix_superradtime_2}. The final result shows:
\begin{equation}
    \frac{da}{d\ln\alpha} = \frac{dt}{dM}\left( \frac{2}{3\alpha\sqrt{3}}B\dot{M}_{\rm ACC} - \frac{\mu}{24}M_S(a-a_{\rm crit})\alpha^7 \right)-2a
    \label{eq:regge_plane_deriv_alpha}
\end{equation}
With the scaling $f_{2,Edd}=\frac{\mu_2}{\mu_1}f_{1,Edd}$ used previously, $\dot{M}_{\rm ACC}$ is constant across $\mu$ at all $\alpha$ (including the accretion events as the time component can be written as $\alpha$ which is shared across $\mu$ in our simulation schemes).

As shown in App. \ref{sec:appendix_superradtime_2}, after the initial superradiance boost $M_S\propto \mu^{-1}$. This means the second term of Eq. \ref{eq:regge_plane_deriv_alpha} is independent of $\mu$. $\dot{M}_S=2M_s\omega_I=\frac{1}{24}M_S(a-a_{\rm crit})\alpha^8\mu$ is also independent of $\mu$ when the $M_S\propto\mu^{-1}$ scaling is applied. As each term of $\dot{M} = \dot{M}_{\rm ACC}-\dot{M}_S$ is independent of $\mu$, $\dot{M}$ will also be independent of $\mu$. This finally leads to every term in Eq. \ref{eq:regge_plane_deriv_alpha} being independent of $\mu$ in our simulation scheme.
This causes the Regge trajectories to align, as the initial $\alpha$ and $a$ are shared before the late boost is applied, since the first superradiance drop has already occurred. 

Note that the derivation in App. \ref{sec:appendix_superradtime_2} cannot be used for boosts before the superradiance drop, as the $M_S\propto\mu^{-1}$ relation does not hold until after the first superradiance boost. 

\subsection{Exclusion Area}
\label{sec:exclusion_area}

\begin{figure}
	\includegraphics[width=\columnwidth]{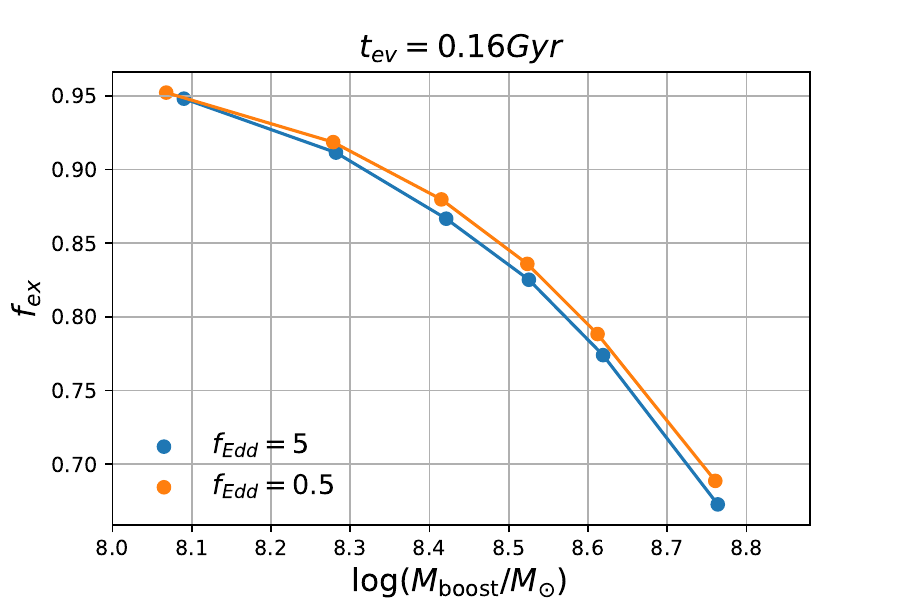}
    \includegraphics[width=\columnwidth]{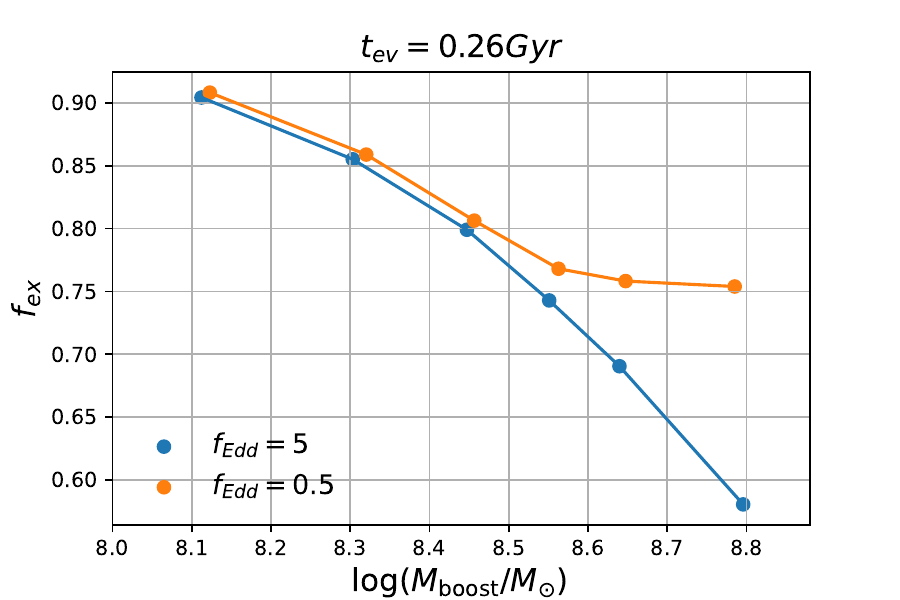}
    \includegraphics[width=\columnwidth]{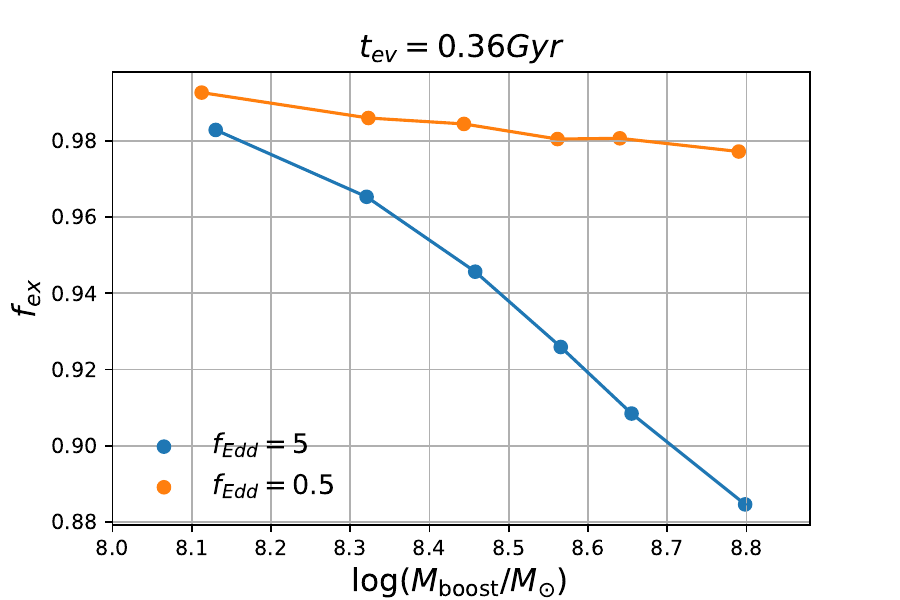}
    \caption{$M_{\rm boost}$ is the total mass accreted during the boost event. As expected, a larger $M_{\rm boost}$ leads to a lower $f_{\rm ex}$. For different $f_{\rm Edd}$ it can be seen that for $t_{\rm ev} = 0.16 \ \rm Gyr$ this has little impact (outlined in Sec. \ref{sec:strong_boost}). In comparison boosts applied at $t_{\rm ev}=0.36Gyr$ follow different trajectories for different $f_{\rm Edd}$ since a higher $f_{\rm Edd}$ has larger impact on exclusion region (outlined in Sec. \ref{sec:strong_boost}). During $t_{\rm ev}=0.26Gyr$, the transition phase occurs where the two $f_{\rm Edd}$ begin to deviate away as the superradiance drop occurs, switching into the late time regime.}
    \label{fig:fex}
\end{figure}

To consolidate our results we consider how the area of the exclusion region changes as a function of the boost parameters. Sections \ref{sec:weak_boost} and \ref{sec:strong_boost} have shown that boosted accretion always reduces the size of the exclusion region. We quantify the impact of accretion boosts by computing $f_{\rm ex}$, the fraction of the area of the exclusion region in the Regge plane with the boost divided by the area of the exclusion region without the boost. 
\begin{equation}
    f_{\rm ex}(x) = \frac{S(x)}{S(\name{noboost})}
\end{equation}
The definition of $S$ is given in Eq. \ref{eq:excl_region_integral} where $x$ denotes a specific simulation scheme.
The smaller $f_{\rm ex}$ the larger the impact of the accretion boost. $f_{\rm ex}= 0$ would be equivalent to a black hole that is always maximally spinning while $f_{\rm ex}= 1$ would be a simulation where the accretion boost has no impact on the exclusion region. In Fig. \ref{fig:fex} we plot $f_{\rm ex}$ as a function of the total mass gained during the accretion boost, $\rm M_{\rm boost}$ for all simulations in Fig. \ref{fig:fedd05_timing} and Fig. \ref{fig:fedd5timing}. 

Firstly, we report a general trend that higher $M_{\rm boost}$ leads to lower $f_{\rm ex}$. This was already noticeable in previous sections where a larger change in black hole mass lead to the exclusion region closing further. Boosts applied before the superradiance drop ($t_{\rm ev} =0.16 \ \rm Gyr$, $t_{\rm ev}=0.26 \rm \ Gyr$) can reduce $f_{\rm ex}$ much more (the top two panels of Fig. \ref{fig:fex} show values of $f_{\rm ex,min}\sim 0.6$) than boosts after the superradiance drop ($t_{\rm ev} =0.36 \rm \ Gyr$, bottom panel of Fig. \ref{fig:fex} has a minimum $f_{\rm ex}\sim 0.88$). 

For different $f_{\rm Edd}$, it can be seen that the behaviours before the superradiance drop are similar. In the $t_{\rm ev} = 0.16 \rm \ Gyr$ case, both trajectories line up closely (explained in Sec. \ref{sec:strong_boost}). Boosts applied after the superradiance drop ($t_{\rm ev} = 0.36 \rm \ Gyr$) show a deviation between the two $f_{\rm Edd}$ trajectories. As outlined in Sec. \ref{sec:strong_boost}, this is because the early time behaviour of the Regge trajectory is dependent on $M_{\rm boost}$ whilst late time behaviour is dependent on $f_{\rm Edd}$. Hence, the higher $f_{\rm Edd}$ gives a low $f_{\rm ex}$ for $t_{\rm ev} = 0.36 \rm Gyr$ since the equilibrium is disrupted more. $t_{\rm ev} = 0.26 \rm \ Gyr$ is the transition period in which the behaviour is initially similar to the $t_{\rm ev} = 0.16 \rm \ Gyr$ case, where both $f_{\rm Edd}$ are closely aligned. The $f_{\rm Edd}=0.5$ scheme at $t_{ev} = 0.26Gyr$ with boosts of $\Delta t_{ev}\ge20Myr$, overlaps with the superradiance drop during the boost itself as can be seen from the middle right panel of Fig. \ref{fig:fedd05_timing}. Therefore, once the drop occurs, longer boosts (higher $M_{\rm boost}$) have a marginal impact on the exclusion region because they do not alter the position of the superradiance drop. The $f_{\rm Edd}=5$ boosts in comparison are completed before the drop, as they have a shorter duration, so they close off more of the exclusion region for the same $M_{\rm boost}$.

The bottom panel of Fig. \ref{fig:fex} confirms that boosts after the superradiance drop have less of an impact on the exclusion region compared to before. In this regime, the strength of the boost rather than $M_{\rm boost}$ matters much more especially for weak boosts where the trend is almost flat with $M_{\rm boost}$. A stronger boost means that the black hole will take longer for $t_s$ to decay and re-equilibrate the system. This allows more angular momentum to be accreted into the black hole causing a stronger deviation in its Regge trajectory. 

\section{Discussion \& Conclusions}
In this paper we presented a study to understand how temporary boosts in mass accretion onto initially highly spinning black holes changes their longterm spin evolution under the influence of superradiance. We particularly studied how parameters of the accretion boost influenced the evolution of the black hole in the Regge plane (the mass-spin or $M- a$ plane). To investigate the effects of variable accretion, accretion boosts were applied before and after the superradiance drop. 

We report that an initially highly spinning black hole will undergo a superradiance "drop" where the spin of the black hole strongly decreases at approximately constant black hole mass, consistent with previous studies \citep{brito,Cardoso_2018}. After the drop, the black hole evolves along the critical spin trajectory in the Regge plane. These two trajectories in the Regge plane define the {\it exclusion region} -- part of the Regge plane that is depopulated of highly spinning black holes due to the superradiance effect. Accretion boosts can change the black hole mass at which the drop occurs, or boost the black hole spin above the critical spin value during the duration of the accretion boost, thus altering the exclusion region. Specifically, we conclude the following:
\begin{itemize}
\setlength\itemsep{0.5em}
\item The evolution of the black hole spin is determined by the relative evolution of two characteristic timescales: the  timescales for angular momentum to be accreted onto the black hole from external accretion $\tacc$ and the timescale for the axion cloud to extract the rotational energy from the black hole $\tsup$. Both timescales continue to evolve until they establish an equilibrium. 
\item 
The initial evolution of the superradiance timescale $\tsup$ is driven by the growing mass of the axion cloud. Once $\tsup$ has fallen sufficiently low to be comparable to the accretion timescale, the superradiance drop occurs. 
\item 
A smaller initial axion cloud mass causes the superradiance drop to occur later as it takes longer for $\tsup$ to decay to the critical level.
\item 
For boosts that begin before the superradiance drop, the later the boost is applied (larger $t_{\rm ev}$ in our nomenclature) the more the exclusion region is closed off. This occurs upto the point of saturation where the drop occurs during the boost itself. Once the drop begins, the shape of the exclusion region is bounded by that of constant accretion at the boosted accretion rate.
\item 
The longer the boost is applied for (larger $\Delta t_{\rm ev}$ in our nomenclature) the more the exclusion region is closed off.
\item 
Boosts after the superradiance drop temporarily deviate the black hole from the critical spin, closing some part of the accretion region. The higher the $f_{\rm Edd}$ of the boost, the greater the increase in spin. 
\item 
Higher axion masses need stronger boosts to have the same impact on the black hole spin evolution\footnote{For two axion particles with masses $\mu_1$ and $\mu_2$, boosts with 
$\Delta t_{\rm ev,1} = \Delta t_{\rm ev,2} \times\,\mu_2/\mu_1$ and $f_{\rm Edd,1} = f_{\rm Edd,2}\times\, \mu_1/\mu_2$ lead to equivalent, but not identical, trajectories on the Regge plane.}. Although exact relation between the axion masses and the strength of the boost is likely impacted by the relativistic corrections.
\end{itemize}

A theoretical framework of analysing timescales was developed in section \ref{sec:timescale_intro}. Using this approach, it was found that the black hole is driven to superradiance through the increase in the axion cloud mass. The evolution after the superradiance drop in spin is due to the decaying $\omega_{\rm I}$ frequency. This in turn keeps the black hole decaying towards the critical spin trajectory (where $\omega_{\rm I}(a_{\rm crit})=0$ which stops the transfer of momentum between the black hole and the axion cloud). A simple Eddington thin disc accretion model was then added to understand the impacts of accretion. Accretion can be seen to drive a black hole to experience the superradiance drop quicker than without accretion. This is because the black hole becomes more massive which increases $\omega_{\rm I}$ and hence increases the rate of cloud growth. Furthermore, an equilibrium state in the timescales of superradiance and accretion emerged after the superradiance drop.

Overall, our work showed that it is the peak accretion rates onto a variably accreting black hole that have the largest impact on the shape of the exclusion region, but that the impact on the accretion region is largest if the boost occurs before the superradiance drop. It was found that variable accretion events before the superradiance drop could have a significant effect in altering the size of the exclusion region for low axion masses ($\mu=10^{-20}$ and lower) for reasonable accretion events with Eddington ratios of $f_{\rm Edd }=5$ lasting for 5 Myr. The timescales during which the black hole spin exceeds that of the corresponding non-boosted simulations exclusion region (order of $10^{7}-10^{8} \rm yr$) is long enough to possibly be observable. The impact of such early boosts closes the accretion region for lower black hole masses. 

We have shown that the superradiance effect is fairly stable against variable accretion effects after the superradiance drop, as for such late accretion events the fraction of the accretion region closed off due to the accretion boost remains small.  Higher axion masses are generally more stable against variable accretion events as unrealistically high, sustained Eddington ratios are required to significantly change the exclusion region. 

One of the caveats of the analysis presented in this paper is exclusion of more realistic self-interaction terms that modify the superradiant states and their time-scales. In particular, self-interaction could delay the spin extraction and also reduce the amount of angular momentum extracted from the black hole. Coupling of the time-varying accretion with more realistic self-interaction models could therefore have strong implications for the exclusion region in the Regge plane. We leave the details of such calculations to future work.

Overall, our results show that the exclusion region remains in existence even for boosts in accretion rate that lead to significant black hole mass growth.One limitation of our study is that we have assumed initially maximally spinning black holes for all our models. A lower initial spin $a_0 << a_{\rm max}$ can also reduce the size of the exclusion region by delaying the superradiance drop \citep{sarmah2025EffectsSuperradianceActive}.
Gravitational wave emission from the axion cloud was also omitted from this study. Adding this dissipative effect would delay the onset of superradiance as the axion cloud would take longer to reach the critical mass. Secondly, boosts that occur after the first superradiance drop would further deviate from the critical spin trajectory due to the interplay between the energy dissipation due to gravitational waves and the accretion boost.
While the boosts tested here have been chosen to be representative of those experienced by massive black holes both in observations and simulations (see Sec. \ref{sec:introduction} for a discussion), our study cannot probe the cumulative effect of continued changes in accretion on the long-term spin evolution of black holes. To do so would require modelling the spin evolution of black holes including the effect of super-radiance based on the black hole's mass accretion histories extracted from cosmological simulations. Such a study would allow us to predict the expected distribution of massive black hole spins for different axion masses. We could also use it to study under which conditions realistic black holes meet the conditions for superradiance drops across cosmic time. We defer this study to future work.

\section*{Acknowledgements}

Beyond the first author, the author list is in alphabetical order. R.S.B. acknowledges support from a UKRI Future Leaders Fellowship (grant code: MR/Y015517/1). For the purpose of open access, the author has applied a Creative Commons Attribution (CC BY) license to any Author Accepted Manuscript version arising from this submission.

\section*{Data Availability}

All data and analysis code used in this work are available from the corresponding author on reasonable request.



\bibliographystyle{mnras}
\bibliography{references,Superradiance_refs_Ricarda}

@article{angles-alcazar2013BLACKHOLEGALAXYCORRELATIONS,
  title = {{{BLACK HOLE-GALAXY CORRELATIONS WITHOUT SELF-REGULATION}}},
  author = {{Angl{\'e}s-Alc{\'a}zar}, Daniel and {\"O}zel, Feryal and Dav{\'e}, Romeel},
  year = {2013},
  month = may,
  volume = {770},
  number = {1},
  pages = {5},
  issn = {0004-637X, 1538-4357},
  doi = {10.1088/0004-637X/770/1/5},
  urldate = {2024-10-16},
  copyright = {http://iopscience.iop.org/info/page/text-and-data-mining},
  journal = {\apj}
}

@article{arvanitaki2011ExploringStringAxiverse,
  title = {Exploring the String Axiverse with Precision Black Hole Physics},
  author = {Arvanitaki, Asimina and Dubovsky, Sergei},
  year = {2011},
  month = feb,
  volume = {83},
  number = {4},
  pages = {044026},
  publisher = {American Physical Society},
  doi = {10.1103/PhysRevD.83.044026},
  urldate = {2025-04-17},
  journal = {\prd}
}

@article{dimatteo2008DirectCosmologicalSimulations,
  title = {Direct {{Cosmological Simulations}} of the {{Growth}} of {{Black Holes}} and {{Galaxies}}},
  author = {Di Matteo, Tiziana and Colberg, J{\"o}rg and Springel, Volker and Hernquist, Lars and Sijacki, Debora},
  year = {2008},
  month = mar,
  volume = {676},
  number = {1},
  pages = {33--53},
  issn = {0004-637X, 1538-4357},
  doi = {10.1086/524921},
  urldate = {2024-09-27},
  language = {en},
  journal = {\apj}
}

@article{dubois2015BlackHoleEvolution,
  title = {Black Hole Evolution - {{I}}. {{Supernova-regulated}} Black Hole Growth},
  author = {Dubois, Yohan and Volonteri, Marta and Silk, Joseph and Devriendt, Julien and Slyz, Adrianne and Teyssier, Romain},
  year = {2015},
  month = sep,
  volume = {452},
  pages = {1502--1518},
  publisher = {OUP},
  issn = {0035-8711},
  doi = {10.1093/mnras/stv1416},
  urldate = {2024-10-29},
  journal = {\mnras}
}

@article{hopkins2006UnifiedMergerdrivenModel,
  title = {A {{Unified}}, {{Merger}}-driven {{Model}} of the {{Origin}} of {{Starbursts}}, {{Quasars}}, the {{Cosmic X}}-{{Ray Background}}, {{Supermassive Black Holes}}, and {{Galaxy Spheroids}}},
  author = {Hopkins, Philip F. and Hernquist, Lars and Cox, Thomas J. and Di Matteo, Tiziana and Robertson, Brant and Springel, Volker},
  year = {2006},
  month = mar,
  volume = {163},
  number = {1},
  pages = {1--49},
  issn = {0067-0049, 1538-4365},
  doi = {10.1086/499298},
  urldate = {2024-09-27},
  language = {en},
  journal = {\apjs}
}

@article{husko2025EffectsSuperEddingtonAccretion,
  title = {The Effects of Super-{{Eddington}} Accretion and Feedback on the Growth of Early Supermassive Black Holes and Galaxies},
  author = {Hu{\v s}ko, Filip and Lacey, Cedric G. and Roper, William J. and Schaye, Joop and Briggs, Jemima Mae and Schaller, Matthieu},
  year = {2025},
  month = mar,
  volume = {537},
  pages = {2559--2578},
  publisher = {OUP},
  issn = {0035-8711},
  doi = {10.1093/mnras/staf146},
  urldate = {2025-02-28},
  journal = {\mnras}
}

@article{lapiner2021CompactiondrivenBlackHole,
  title = {Compaction-Driven Black Hole Growth},
  author = {Lapiner, Sharon and Dekel, Avishai and Dubois, Yohan},
  year = {2021},
  month = jul,
  volume = {505},
  pages = {172--190},
  publisher = {OUP},
  issn = {0035-8711},
  doi = {10.1093/mnras/stab1205},
  urldate = {2024-10-29},
  journal = {\mnras}
}

@article{lupi2024SustainedSuperEddingtonAccretion,
  title = {Sustained Super-{{Eddington}} Accretion in High-Redshift Quasars},
  author = {Lupi, Alessandro and Quadri, Giada and Volonteri, Marta and Colpi, Monica and Regan, John A.},
  year = {2024},
  month = jun,
  volume = {686},
  pages = {A256},
  issn = {0004-6361, 1432-0746},
  doi = {10.1051/0004-6361/202348788},
  urldate = {2025-03-10},
  copyright = {https://creativecommons.org/licenses/by/4.0},
  journal = {\aap}
}

@article{massonneau2023SuperEddingtonGrowth,
  title = {How the Super-{{Eddington}} Regime Regulates Black Hole Growth in High-Redshift Galaxies},
  author = {Massonneau, Warren and Volonteri, Marta and Dubois, Yohan and Beckmann, Ricarda S.},
  year = {2023},
  month = feb,
  volume = {670},
  pages = {A180},
  doi = {10.1051/0004-6361/202243170},
  journal = {\aap}
}

@article{mcalpine2018RapidGrowthPhase,
  title = {The Rapid Growth Phase of Supermassive Black Holes},
  author = {McAlpine, Stuart and Bower, Richard G. and Rosario, David J. and Crain, Robert A. and Schaye, Joop and Theuns, Tom},
  year = {2018},
  month = dec,
  volume = {481},
  pages = {3118--3128},
  publisher = {OUP},
  issn = {0035-8711},
  doi = {10.1093/mnras/sty2489},
  urldate = {2024-10-29},
  journal = {\mnras}
}

@article{mclure2004CosmologicalEvolutionQuasar,
  title = {The Cosmological Evolution of Quasar Black Hole Masses},
  author = {McLure, Ross J. and Dunlop, James S.},
  year = {2004},
  month = aug,
  volume = {352},
  number = {4},
  pages = {1390--1404},
  issn = {00358711, 13652966},
  doi = {10.1111/j.1365-2966.2004.08034.x},
  urldate = {2024-06-07},
  language = {en},
  journal = {\mnras}
}

@article{prieto2017HowAGNSN,
  title = {How {{AGN}} and {{SN Feedback Affect Mass Transport}} and {{Black Hole Growth}} in {{High-redshift Galaxies}}},
  author = {Prieto, Joaquin and Escala, Andr{\'e}s and Volonteri, Marta and Dubois, Yohan},
  year = {2017},
  month = feb,
  volume = {836},
  pages = {216},
  publisher = {IOP},
  issn = {0004-637X},
  doi = {10.3847/1538-4357/aa5be5},
  urldate = {2024-10-25},
  journal = {\apj}
}

@article{regan2019SuperEddingtonAccretionFeedback,
  title = {Super-{{Eddington}} Accretion and Feedback from the First Massive Seed Black Holes},
  author = {Regan, John A. and Downes, Turlough P. and Volonteri, Marta and Beckmann, Ricarda and Lupi, Alessandro and Trebitsch, Maxime and Dubois, Yohan},
  year = {2019},
  month = jul,
  volume = {486},
  number = {3},
  pages = {3892--3906},
  doi = {10.1093/mnras/stz1045},
  journal = {\mnras}
}

@article{sarmah2025EffectsSuperradianceActive,
  title = {Effects of {{Superradiance}} in {{Active Galactic Nuclei}}},
  author = {Sarmah, Priyanka and Verma, Himanshu and Cheung, Kingman and Silk, Joseph},
  year = {2025},
  month = mar,
  volume = {538},
  number = {2},
  eprint = {2404.09955},
  primaryclass = {astro-ph},
  pages = {943--962},
  issn = {0035-8711, 1365-2966},
  doi = {10.1093/mnras/staf326},
  urldate = {2025-04-17},
  archiveprefix = {arXiv},
  journal = {\mnras}
}

@article{sotan1982MassesQuasars,
  title = {Masses of Quasars},
  author = {Soltan, A.},
  year = {1982},
  month = sep,
  volume = {200},
  number = {1},
  pages = {115--122},
  issn = {0035-8711, 1365-2966},
  doi = {10.1093/mnras/200.1.115},
  urldate = {2024-11-01},
  language = {en},
  journal = {\mnras}
}

@article{trakhtenbrot2011BLACKHOLEMASS,
  title = {{{BLACK HOLE MASS AND GROWTH RATE AT}} {\emph{z}} {$\simeq$} 4.8: {{A SHORT EPISODE OF FAST GROWTH FOLLOWED BY SHORT DUTY CYCLE ACTIVITY}}},
  shorttitle = {{{BLACK HOLE MASS AND GROWTH RATE AT}} {\emph{z}} {$\simeq$} 4.8},
  author = {Trakhtenbrot, Benny and Netzer, Hagai and Lira, Paulina and Shemmer, Ohad},
  year = {2011},
  month = mar,
  volume = {730},
  number = {1},
  pages = {7},
  issn = {0004-637X, 1538-4357},
  doi = {10.1088/0004-637X/730/1/7},
  urldate = {2024-09-13},
  journal = {\apj}
}

@article{yu2002ObservationalConstraintsGrowth,
  title = {Observational Constraints on Growth of Massive Black Holes},
  author = {Yu, Q. and Tremaine, S.},
  year = {2002},
  month = oct,
  volume = {335},
  number = {4},
  pages = {965--976},
  issn = {0035-8711, 1365-2966},
  doi = {10.1046/j.1365-8711.2002.05532.x},
  urldate = {2024-09-11},
  language = {en},
  journal = {\mnras}
}

@article{bardeenKerrMetricBlack1970,
  title = {Kerr {{Metric Black Holes}}},
  author = {Bardeen, James M.},
  year = {1970},
  month = apr,
  journal = {Nature},
  volume = {226},
  number = {5240},
  pages = {64--65},
  issn = {0028-0836, 1476-4687},
  doi = {10.1038/226064a0},
  urldate = {2024-06-21},
  copyright = {http://www.springer.com/tdm},
  langid = {english}
}

@article{paper1,
  title = {Black hole spin constraints on the mass spectrum and number of axionlike fields},
  author = {Stott, Matthew J. and Marsh, David J. E.},
  journal = {\prd},
  volume = {98},
  issue = {8},
  pages = {083006},
  numpages = {24},
  year = {2018},
  month = {Oct},
  publisher = {American Physical Society},
  doi = {10.1103/PhysRevD.98.083006},
  url = {https://link.aps.org/doi/10.1103/PhysRevD.98.083006}
}

@ARTICLE{brito,
       author = {{Brito}, Richard and {Cardoso}, Vitor and {Pani}, Paolo},
        title = "{Black holes as particle detectors: evolution of superradiant instabilities}",
      journal = {Classical and Quantum Gravity},
         year = 2015,
        month = jul,
       volume = {32},
       number = {13},
          eid = {134001},
        pages = {134001},
          doi = {10.1088/0264-9381/32/13/134001},
archivePrefix = {arXiv},
       eprint = {1411.0686},
 primaryClass = {gr-qc},
       adsurl = {https://ui.adsabs.harvard.edu/abs/2015CQGra..32m4001B}
}

@article{Baumann1,
       author = {{Baumann}, Daniel and {Chia}, Horng Sheng and {Stout}, John and {ter Haar}, Lotte},
        title = "{The spectra of gravitational atoms}",
      journal = {\jcap},
         year = 2019,
        month = dec,
       volume = {2019},
       number = {12},
          eid = {006},
        pages = {006},
          doi = {10.1088/1475-7516/2019/12/006},
archivePrefix = {arXiv},
       eprint = {1908.10370},
 primaryClass = {gr-qc},
       adsurl = {https://ui.adsabs.harvard.edu/abs/2019JCAP...12..006B}
}

@article{Baumann2,
       author = {{Baumann}, Daniel and {Chia}, Horng Sheng and {Porto}, Rafael A.},
        title = "{Probing ultralight bosons with binary black holes}",
      journal = {\prd},
         year = 2019,
        month = feb,
       volume = {99},
       number = {4},
          eid = {044001},
        pages = {044001},
          doi = {10.1103/PhysRevD.99.044001},
archivePrefix = {arXiv},
       eprint = {1804.03208},
 primaryClass = {gr-qc},
       adsurl = {https://ui.adsabs.harvard.edu/abs/2019PhRvD..99d4001B}
}

@article{Hui,
       author = {{Hui}, Lam and {Law}, Y.~T. Albert and {Santoni}, Luca and {Sun}, Guanhao and {Tomaselli}, Giovanni Maria and {Trincherini}, Enrico},
        title = "{Black hole superradiance with dark matter accretion}",
      journal = {\prd},
         year = 2023,
        month = may,
       volume = {107},
       number = {10},
          eid = {104018},
        pages = {104018},
          doi = {10.1103/PhysRevD.107.104018},
archivePrefix = {arXiv},
       eprint = {2208.06408},
 primaryClass = {gr-qc},
       adsurl = {https://ui.adsabs.harvard.edu/abs/2023PhRvD.107j4018H}
}

@article{Ricarda,
       author = {{Beckmann}, R.~S. and {Devriendt}, J. and {Slyz}, A.},
        title = "{Zooming in on supermassive black holes: how resolving their gas cloud host renders their accretion episodic}",
      journal = {\mnras},
         year = 2019,
        month = mar,
       volume = {483},
       number = {3},
        pages = {3488-3509},
          doi = {10.1093/mnras/sty2890},
archivePrefix = {arXiv},
       eprint = {1810.01649},
 primaryClass = {astro-ph.GA},
       adsurl = {https://ui.adsabs.harvard.edu/abs/2019MNRAS.483.3488B}
}

@article{kip,
       author = {{Thorne}, Kip S.},
        title = "{Disk-Accretion onto a Black Hole. II. Evolution of the Hole}",
      journal = {\apj},
         year = 1974,
        month = jul,
       volume = {191},
        pages = {507-520},
          doi = {10.1086/152991},
       adsurl = {https://ui.adsabs.harvard.edu/abs/1974ApJ...191..507T}
}

@article{zeldovich,
       author = {{Zel'Dovich}, Ya. B.},
        title = "{Generation of Waves by a Rotating Body}",
      journal = {Soviet Journal of Experimental and Theoretical Physics Letters},
         year = 1971,
        month = aug,
       volume = {14},
        pages = {180},
       adsurl = {https://ui.adsabs.harvard.edu/abs/1971JETPL..14..180Z}
}

@article{Cardoso_2018,
       author = {{Cardoso}, Vitor and {Dias}, {\'O}scar J.~C. and {Hartnett}, Gavin S. and {Middleton}, Matthew and {Pani}, Paolo and {Santos}, Jorge E.},
        title = "{Constraining the mass of dark photons and axion-like particles through black-hole superradiance}",
      journal = {\jcap},
         year = 2018,
        month = mar,
       volume = {2018},
       number = {3},
          eid = {043},
        pages = {043},
          doi = {10.1088/1475-7516/2018/03/043},
archivePrefix = {arXiv},
       eprint = {1801.01420},
 primaryClass = {gr-qc},
       adsurl = {https://ui.adsabs.harvard.edu/abs/2018JCAP...03..043C}
}

@article{Witte:2024drg,
       author = {{Witte}, Samuel J. and {Mummery}, Andrew},
        title = "{Stepping up superradiance constraints on axions}",
      journal = {\prd},
         year = 2025,
        month = apr,
       volume = {111},
       number = {8},
          eid = {083044},
        pages = {083044},
          doi = {10.1103/PhysRevD.111.083044},
archivePrefix = {arXiv},
       eprint = {2412.03655},
 primaryClass = {hep-ph},
       adsurl = {https://ui.adsabs.harvard.edu/abs/2025PhRvD.111h3044W}
}

@article{TaillteMay,
       author = {{May}, Taillte and {East}, William E. and {Siemonsen}, Nils},
        title = "{Self-gravity effects of ultralight boson clouds formed by black hole superradiance}",
      journal = {\prd},
         year = 2025,
        month = feb,
       volume = {111},
       number = {4},
          eid = {044062},
        pages = {044062},
          doi = {10.1103/PhysRevD.111.044062},
archivePrefix = {arXiv},
       eprint = {2410.21442},
 primaryClass = {gr-qc},
       adsurl = {https://ui.adsabs.harvard.edu/abs/2025PhRvD.111d4062M}
}

@article{chargedbh,
       author = {{Li}, Qian and {Wang}, Qianchuan and {Jia}, Junji},
        title = "{Absorption and scattering of charged scalar waves by charged Horndeski black hole}",
      journal = {\prd},
         year = 2025,
        month = jan,
       volume = {111},
       number = {2},
          eid = {024059},
        pages = {024059},
          doi = {10.1103/PhysRevD.111.024059},
archivePrefix = {arXiv},
       eprint = {2411.02987},
 primaryClass = {gr-qc},
       adsurl = {https://ui.adsabs.harvard.edu/abs/2025PhRvD.111b4059L}
}

@article{PhysRevD.100.044051,
  title = {Superradiant scattering by a black hole binary},
  author = {Wong, Leong Khim},
  journal = {\prd},
  volume = {100},
  issue = {4},
  pages = {044051},
  numpages = {8},
  year = {2019},
  month = {Aug},
  publisher = {American Physical Society},
  doi = {10.1103/PhysRevD.100.044051},
  url = {https://link.aps.org/doi/10.1103/PhysRevD.100.044051}
}

@article{Fan:2023jjj,
    author = "Fan, Kaiyuan and Tong, Xi and Wang, Yi and Zhu, Hui-Yu",
    title = "{Modulating binary dynamics via the termination of black hole superradiance}",
    eprint = "2311.17013",
    archivePrefix = "arXiv",
    primaryClass = "gr-qc",
    doi = "10.1103/PhysRevD.109.024059",
    journal = "\prd",
    volume = "109",
    number = "2",
    pages = "024059",
    year = "2024"
}

@article{Starobinsky,
       author = {{Starobinski{\v{i}}}, A.~A.},
        title = "{Amplification of waves during reflection from a rotating ``black hole''}",
      journal = {Soviet Journal of Experimental and Theoretical Physics},
         year = 1973,
        month = jul,
       volume = {37},
        pages = {28},
       adsurl = {https://ui.adsabs.harvard.edu/abs/1973JETP...37...28S}
}

@article{bardeen,
       author = {{Bardeen}, James M. and {Press}, William H. and {Teukolsky}, Saul A.},
        title = "{Rotating Black Holes: Locally Nonrotating Frames, Energy Extraction, and Scalar Synchrotron Radiation}",
      journal = {\apj},
         year = 1972,
        month = dec,
       volume = {178},
        pages = {347-370},
          doi = {10.1086/151796},
       adsurl = {https://ui.adsabs.harvard.edu/abs/1972ApJ...178..347B}
}

@article{PhysRevD.22.2323,
  title = {Klein-Gordon equation and rotating black holes},
  author = {Detweiler, Steven},
  journal = {\prd},
  volume = {22},
  issue = {10},
  pages = {2323--2326},
  numpages = {0},
  year = {1980},
  month = {Nov},
  publisher = {American Physical Society},
  doi = {10.1103/PhysRevD.22.2323},
  url = {https://link.aps.org/doi/10.1103/PhysRevD.22.2323}
}

@article{PhysRevD.95.043001,
  title = {Black hole mergers and the QCD axion at Advanced LIGO},
  author = {Arvanitaki, Asimina and Baryakhtar, Masha and Dimopoulos, Savas and Dubovsky, Sergei and Lasenby, Robert},
  journal = {\prd},
  volume = {95},
  issue = {4},
  pages = {043001},
  numpages = {6},
  year = {2017},
  month = {Feb},
  publisher = {American Physical Society},
  doi = {10.1103/PhysRevD.95.043001},
  url = {https://link.aps.org/doi/10.1103/PhysRevD.95.043001}
}

@article{PhysRevD.103.095019,
  title = {Black hole superradiance of self-interacting scalar fields},
  author = {Baryakhtar, Masha and Galanis, Marios and Lasenby, Robert and Simon, Olivier},
  journal = {\prd},
  volume = {103},
  issue = {9},
  pages = {095019},
  numpages = {61},
  year = {2021},
  month = {May},
  publisher = {American Physical Society},
  doi = {10.1103/PhysRevD.103.095019},
  url = {https://link.aps.org/doi/10.1103/PhysRevD.103.095019}
}

@article{Saha_2022,
   title={Evolution of primordial black holes in an adiabatic FLRW universe with gravitational particle creation},
   volume={54},
   ISSN={1572-9532},
   url={http://dx.doi.org/10.1007/s10714-022-03010-6},
   DOI={10.1007/s10714-022-03010-6},
   number={10},
   journal={General Relativity and Gravitation},
   publisher={Springer Science and Business Media LLC},
   author={Saha, Subhajit and Mamon, Abdulla Al and Saha, Somnath},
   year={2022},
   month=oct }

@article{Davoudiasl_2019,
   title={Ultralight Boson Dark Matter and Event Horizon Telescope Observations of M87},
   volume={123},
   ISSN={1079-7114},
   url={http://dx.doi.org/10.1103/PhysRevLett.123.021102},
   DOI={10.1103/physrevlett.123.021102},
   number={2},
   journal={\prl},
   publisher={American Physical Society (APS)},
   author={Davoudiasl, Hooman and Denton, Peter B.},
   year={2019},
   month=jul }

@article{PhysRevD.87.043513,
  title = {Superradiant instabilities in astrophysical systems},
  author = {Witek, Helvi and Cardoso, Vitor and Ishibashi, Akihiro and Sperhake, Ulrich},
  journal = {\prd},
  volume = {87},
  issue = {4},
  pages = {043513},
  numpages = {25},
  year = {2013},
  month = {Feb},
  publisher = {American Physical Society},
  doi = {10.1103/PhysRevD.87.043513},
  url = {https://link.aps.org/doi/10.1103/PhysRevD.87.043513}
}




\appendix

\section{Time of Superradiance drop dependence on $\mu$}
\label{sec:appendix_superradtime}
To prove the alignment of Regge trajectories for boosts before the superradiance drop, we must show that the $\alpha$ at which the drop occurs at is independent of $\mu$. This is first done through deriving the time at which the superradiance drop occurs for the constant accretion case (constant $f_{Edd}$). This can then be trivially extended to a boosted situation as the boost can be considered as another regime of constant accretion.  

The following derivation uses natural units ($c=1,G=1,\hbar=1$). As outlined in Sec. \ref{sec:timescale_intro}, initial $\omega_{\rm I}$ is assumed constant before the superradiance drop compared to axion cloud growth.
 
Firstly, from Eq. \ref{eq:tsup} we have $\log{\tsup}\sim \log{\dfrac{M^2}{M_{\rm S}}} - \log{\omega_{\rm I}} + \log{\frac{a\mu}{2}}$. Using $\omega_{\rm I} = \frac{1}{48}(a-a_{\rm crit})\alpha^8\mu$ from Eq. \ref{eq:freqspin} and substituting explicit $M$ dependencies with $\frac{\alpha}{\mu}$ we get: 
\begin{equation}
\label{eq:app_a_highaxtsup}
    \log{\tsup} = \log\left(\frac{24a}{\alpha^6M_{\rm S}\mu^2(a-a_{\rm crit})} \right)
\end{equation}

Secondly, before the superradiance drop, the decay of $\tsup$ is driven by the growth of the axion cloud rather than $\omega_I$ as explained in Sec. \ref{sec:timescale_intro}. This means $\frac{d\log M_{\rm S}}{dt} >> \frac{d\log \omega_{\rm I}}{dt}$ and $\frac{d\log M_{\rm S}}{dt}>>\frac{d\log M}{dt}$ which is also established in Sec. \ref{sec:timescale_intro}. From this we can see that the derivative of $\log{\tsup}\sim \log{M^2} -\log{M_{\rm S}} - \log{\omega_{\rm I}} + \log{\frac{a\mu}{2}}$, is $\dfrac{d\log \tsup}{dt} \approx -\dfrac{d\log M_{\rm S}}{dt} = -2\omega_{\rm I}$.  
Once again using $\omega_{\rm I} = \frac{1}{48}(a-a_{\rm crit})\alpha^8\mu$ from Eq. \ref{eq:freqspin} gives
\begin{equation}
\label{eq:highaxderivtsup}
    \dfrac{d\log\tsup}{dt} \approx -\dfrac{\dot{M}_S}{M_{\rm S}}=-\frac{1}{24}(a-a_{\rm crit})\alpha^8\mu.
\end{equation}
For current purposes, we assume $\frac{d\log\tsup}{dt}$ to be relatively constant before the superradiance drop which, is correct to the first order.

Moving onto $\tacc$, from Eq. \ref{eq:tacc} we can write $\tacc = \dfrac{Ma}{\dot{M}_{\rm acc}}$ and using $\dot{M}_{\rm acc}=f_{\rm Edd}M$ gives 
\begin{equation}
\label{eq:app_a_highaxtacc}
    \log\tacc = \log\frac{a}{f_{\rm Edd}}
\end{equation}

Assuming the rate of decay of $\tsup$ is constant in log space, the time between the beginning of the evolution and the superradiance drop (here called $t_s$) is given by 
\begin{equation}
    t_s = \frac{\log \tacc - \log\tsup}{\frac{d\log\tsup}{dt}}
    \label{eq:taus_temp1}
\end{equation}
Where all quantities ($a$, $a_{crit}$, $M_{S}$ and $\alpha$) are evaluated at time $t=0$ as we assume constant decay in $\tsup$.

The only exception might be the black hole spin $a$, which we assume to be maximal for the purpose of this discussion. This is justified because for our simulations $\tacc << \tsup$ initially, which means the black hole will always spin up maximally long before the superradiance drop.  Plugging Eq. \ref{eq:app_a_highaxtsup}, \ref{eq:highaxderivtsup} and \ref{eq:app_a_highaxtacc} into Eq. \ref{eq:taus_temp1} gives
\begin{equation}
    t_s = \frac{24}{(a-a_{\rm crit})\alpha^8\mu}\log\left(\frac{24f_{\rm Edd}}{\alpha^6M_{\rm S}\mu^2(a-a_{\rm crit})}\right)
\end{equation}
where all terms are constant and evaluated at $t=0$. 

So for different axion masses of $\mu_1$ and $\mu_2$ we expect 
\begin{equation}
    t_{s,1} = \frac{A_2}{A_1}\left[t_{s,2} -\frac{24}{A_2}\log\left(\frac{\alpha_1^6M_{1,S}\mu_1^2f_{2,\rm Edd}(a_1-a_{1,\rm crit})}{\alpha_2^6M_{2,S}\mu_2^2f_{1,\rm Edd}(a_2-a_{2,\rm crit})}\right)\right]
    \label{eq:axion_mass_delay_appendix}
\end{equation}
where
\begin{equation}
    A_i=(a_i-a_{i,\rm crit})\alpha_i^8\mu_i.
\end{equation}
From this, the $\alpha$ at which the superradiance drop occurs (denoted as $\alpha_S$) is given by $\alpha_S = \alpha_{t=0}\exp{(f_{\rm Edd}t_s)}$ as $\frac{dM}{dt}>>\frac{dM_S}{dt}$ before the superradiance drop\footnote{ Note that this is not in contradiction to the previous result of $\frac{d\log M_S}{dt}>>\frac{d\log M}{dt}$}. Considering shared initial $\alpha$, $(a-a_{\rm crit})$ and $M_{\rm S}$ as in our simulation scheme, we see that Eq. \ref{eq:axion_mass_delay_appendix} has a scaling of $\frac{\mu_2}{\mu_1}$ with a time delay as the second term. This time delay is much smaller than the leading $t_{s,2}$ term. This means $t_{s,1}\approx\frac{\mu_2}{\mu_1}t_{s,2}$. Therefore, as $f_{1,\rm Edd}=\frac{\mu_1}{\mu_2}f_{2,\rm Edd}$ in our simulation scheme, $\alpha_{\rm S}$ is the same across $\mu_1$ and $\mu_2$ causing the Regge trajectories to align. For a boosted case, the boost can be considered as another regime of constant accretion, so the time scalings hold. The only caveat is that from Eq. \ref{eq:axcloud_variation}, $M_{1,S}\ne M_{2,S}$ after the boosts. However, this only contributes to the time delay term, which is usually small. This means that the $t_{s,1}\approx\frac{\mu_2}{\mu_1}t_{s,2}$ scaling will still largely hold.

\section{Regge Plane evolution dependance on $\mu$}
\label{sec:appendix_superradtime_2}
To prove the alignment of the trajectories in the Regge space, we must show that $\frac{da}{d\ln\alpha}$ is independent of $\mu$. Therefore, if initial conditions ($a$, $\alpha$ and $M_S$) are shared, then the Regge trajectories will align. Using natural units ($c=1$, $G=1$, $\hbar = 1$), the mathematical derivation of this is given below.

First, the derivative of $a$ with $\ln\alpha$ is manipulated using chain rule to give the following:
\begin{equation}
    \frac{da}{d\ln\alpha} = \alpha\frac{da}{d\alpha} =\frac{\alpha}{\mu}\frac{da}{dM}
\end{equation}
Using the definition of spin ($a=J/M^2$) we take the derivatives with respect to $M$ give
\begin{equation}
    \frac{\alpha}{\mu}\frac{da}{dM} =\frac{\alpha}{\mu}\left(\frac{1}{M^2}\frac{dJ}{dM}-\frac{2J}{M^3}\right) 
\end{equation}
$a$ is substituted to remove any $J$ terms and any explicit $M$ dependencies are replaced with $\frac{\alpha}{\mu}$.
\begin{equation}
    \frac{\alpha}{\mu}\left(\frac{1}{M^2}\frac{dJ}{dM}-\frac{2J}{M^3}\right) = \frac{\mu}{\alpha}\frac{dJ}{dM}-2a
    \label{eq:appendix_mass_deriv_regge}
\end{equation}
The derivative of angular momentum ($J$) with $M$ can be converted to a time derivative $\frac{dJ}{dM} = \frac{dt}{dM}\frac{dJ}{dt}$. This is used in Eq. \ref{eq:appendix_mass_deriv_regge} to give:
\begin{equation}
    \frac{da}{d\ln\alpha}=\frac{\mu}{\alpha}\frac{dt}{dM}\frac{dJ}{dt}-2a
    \label{eq:regge_deriv_appendix}
\end{equation}
$\frac{dJ}{dt}$ is the time derivative of the BH angular momentum, which is described in Eq. \ref{eq:AMconserveequation} as $\dot{J} = \dot{J}_{\rm ACC} - \frac{\dot{M}_s}{\mu}$. We will analyze the terms $\dot{J}_{\rm ACC}$ and $\frac{\dot{M}_s}{\mu}$ separately. $\dot{J}_{\rm ACC}$ is represented in Eq. \ref{eq:Jaccterm} as $\dot{J}_{\rm ACC}=\frac{2M}{3\sqrt{3}}B\dot{M}_{\rm ACC}$. $B$ is defined from Eq. \ref{eq:Jaccterm} as 
\begin{equation}
    B=\frac{2G}{3c\sqrt{3}}\frac{1+2\sqrt{\frac{3c^{2}r_{\rm ISCO}}{GM}-2}}{\sqrt{1-\frac{2GM}{3c^{2}r_{\rm ISCO}}}}
    \label{eq:Bdef_appendix_b}
\end{equation}
From \citep{bardeen}, the innermost stable circular orbit is given by
\begin{equation}
r_{\rm ISCO} = \frac{GM}{c^2} \left[ 3 + Z_2(a) \pm \sqrt{(3 - Z_1(a))(3 + Z_1(a) + 2 Z_2(a))} \right],
\end{equation}
where
\begin{equation}
\begin{split}
Z_1(a) &= 1 + (1 - a^2)^{1/3} 
          \Big[ (1 + a)^{1/3} + (1 - a)^{1/3} \Big], \\
Z_2(a) &= \sqrt{3 a^2 + Z_1(a)^2}.
\end{split}
\end{equation}
Since $Z_1$ and $Z_2$ depend explicitly on the spin parameter $a$, all occurrences of $\frac{r_{\rm ISCO}}{M}$ in $B$ of Eq.~\ref{eq:Bdef_appendix_b} are functions of $a$ only, as the mass $M$ cancels. Hence, $B$ itself is solely a function of $a$, with all other dependencies (including $\mu$) removed. 
Next, $\frac{\dot{M}_S}{\mu}$ can be rewritten as $\frac{2M_S\omega_I}{\mu}$ by substituting $\dot{M}_S$ through Eq. \ref{eq:dotMcloud}. Using $\omega_{\rm I} = \frac{1}{48}(a-a_{\rm crit})\alpha^8\mu$ from Eq. \ref{eq:freqspin} gives:
\begin{equation}
    \frac{\dot{M}_s}{\mu}=\frac{1}{24}M_s(a-a_{\rm crit})\alpha^8
\end{equation}
Using our definitions of $\dot{J}_{\rm ACC}$ and $\frac{\dot{M}_S}{\mu}$, we can rewrite $\frac{dJ}{dt}$ through Eq. \ref{eq:regge_deriv_appendix} as:
\begin{equation}
    \frac{dJ}{dt} = \frac{2\alpha}{3\mu\sqrt{3}}B\dot{M}_{\rm ACC}-\frac{1}{24}M_S(a-a_{\rm crit})\alpha^8
\end{equation}
This can then be used to substitute $\frac{dJ}{dt}$ in Eq. \ref{eq:regge_deriv_appendix} to finally give:
\begin{equation}
    \frac{da}{d\ln\alpha} = \frac{dt}{dM}\left( \frac{2}{3\sqrt{3}}B\dot{M}_{\rm ACC} - \frac{\mu}{24}M_S(a-a_{\rm crit})\alpha^7 \right)-2a
    \label{eq:app_b_final_eq}
\end{equation}
To prove Eq. \ref{eq:app_b_final_eq} is independent of $\mu$ we must consider how $M_S$ varies for different $\mu$. 
From Sec. \ref{sec:timescale_intro} we know that before the superradiance drop, $M_S$ increases to drive $\tsup$ to $\tacc$, causing the superradiance drop. After the superradiance drop, $M_S$ is relatively constant. In our simulation scheme, the initial $M_S$ values are equal across $\mu$. As the growth of $M_S$ is what drives $\tsup$ to $\tacc$, through Eq. \ref{eq:superradtimescale}: $\log M_S\propto \log \tsup(t=0) - \log\tacc$ where $\tsup(t=0)$ is the superradiance timescale evaluated at the beginning of the simulation. Using Eq. \ref{eq:app_a_highaxtsup} and Eq. \ref{eq:app_a_highaxtacc} we get that $M_S$ at the point of the superradiance drop is
\begin{equation}
    M_S\propto \frac{24f_{\rm Edd}}{\mu^2\alpha^6M_{S}(a-a_{\rm crit})}
\end{equation}
Where $\alpha$, $M_S$, $a$ and $a_{crit}$ are all evaluated at $t=0$.
In our simulation scheme $f_{\rm Edd} \propto \mu$ pre-boost. Therefore at the point of the superradiance drop, 
\begin{equation}
    M_S\propto \mu^{-1}
\end{equation}
where we have ignored any shared constant terms (such as $\alpha$, $M_{S}$ and $(a_-a_{\rm crit})$ that are shared between $\mu$ in our simulation setup when evaluated at $t=0$). 

 To prove that the relation $M_S\propto \mu^{-1}$ holds at later times after the superradiance drop, we must show that $\frac{dM_S}{dt}$ is also independent of $\mu$. Using Eq. \ref{eq:dotMcloud} we see:
\begin{equation}
    \dot{M}_S = 2M_S\omega_I=\frac{1}{24}M_S(a-a_{\rm crit})\alpha^8\mu
\end{equation}
 Since $M_S\propto\mu^{-1}$ right after the initial superradiance drop, the dependence on $\mu$ is removed. Therefore, a $M_S\propto \mu^{-1}$ scaling remains throughout the evolution of the black hole after the initial superradiance drop. Referring to Eq. \ref{eq:app_b_final_eq}, the second term within the brackets, involving $M_S$, cancels with the factor of $\mu$, eliminating any dependence on $\mu$. Furthermore, the mass evolution equation, $\frac{dM}{dt} = \dot{M}_{\rm ACC} - \dot{M}_S$, is likewise independent of $\mu$, since both $\dot{M}_{\rm ACC}$ and $\dot{M}_S$ are independent of it. These relations will continue to exist for any subsequent boost after the initial superradiance drop.
 As a result, Eq. \ref{eq:app_b_final_eq} is entirely independent of $\mu$, and black holes will follow identical trajectories in the Regge plane even for boosts after the initial superradiance drop, provided the appropriate scaling relations are applied.
 

\bsp	
\label{lastpage}
\end{document}